\documentclass[journal,10pt]{IEEEtran}
\usepackage{amsmath,amsfonts}
\usepackage{threeparttable}
\usepackage{graphicx}
\usepackage{siunitx}
\usepackage{multirow}
\usepackage{makecell}
\usepackage{array}
\usepackage{amsmath}
\usepackage{amssymb}
\usepackage{epstopdf}
\usepackage{float}
\usepackage{amssymb}
\usepackage{nomencl}
\usepackage{enumerate}
\usepackage[linesnumbered,ruled,vlined]{algorithm2e}
\usepackage{subfigure}
\usepackage{bm}
\usepackage{amsthm}

\usepackage{multirow}
\usepackage{array}
\usepackage[caption=false,font=normalsize,labelfont=sf,textfont=sf]{subfig}
\usepackage{textcomp}
\usepackage{stfloats}
\usepackage{url}
\usepackage{xcolor} 
\usepackage{verbatim}
\usepackage{graphicx}
\usepackage{cite}
\usepackage{booktabs}
\newcommand{\RNum}[1]{\uppercase\expandafter{\romannumeral #1\relax}}
\usepackage[english]{babel}
\newtheorem{definition}{Definition}
\newtheorem{theorem}{{Theorem}}

\begin{document}

\title{Multilevel Graph  Reinforcement Learning for Consistent Cognitive Decision-making in Heterogeneous Mixed Autonomy
% Cognitive Multitasking in Heterogeneous Mixed Autonomy: A  Multi-level Graph Reinforcement Learning Approach

% A Spatio-Temporal Interaction-Enhanced Decision-Making Strategy for Connected Autonomous Vehicles in Heterogeneous Mixed Autonomy
}
% Rate GQN: A Decision-making Method Based on  Graph Convolution Deep Reinforcement Learning for Connected and Automated Vehicles in Spatio-temporal Interactive Scenarios
% Scenarios with space-time entanglement
\author{
Xin Gao$^{1}$,~\IEEEmembership{Graduate Student Member,~IEEE}, Zhaoyang Ma$^{1}$, Xueyuan Li*,   Xiaoqiang Meng,  Zirui Li,~\IEEEmembership{Graduate Student Member,~IEEE}
\thanks{
$^{1}$ Xin Gao and Zhaoyang Ma contribute equally to this work.

*Corresponding author: Xueyuan Li 

Xin Gao, Xueyuan Li, Xiaoqiang Meng and Zirui Li  are with the School of Mechanical Engineering, Beijing Institute of Technology, Beijing, China. (E-mails: x.gao@bit.edu.cn; lixueyuan@bit.edu.cn;  3120220377@bit.edu.cn;  z.li@bit.edu.cn.

Zhaoyang Ma is with the School of Computer and Information Technology, Beijing Jiaotong University, Beijing, China. (E-mails:zhy.ma@bjtu.edu.cn)

Zirui Li is also with the Chair of Trafﬁc Process Automation, ”FriedrichList” Faculty of Transport and Trafﬁc Sciences, TU Dresden, Germany.

}% <-this % stops a space
}

%{Shell \MakeLowercase{\textit{et al.}}: Bare Demo of IEEEtran.cls for Journals}
\markboth{Journal of \LaTeX\ Class Files,~Vol.~14, No.~8, August~2021}%
{Shell \MakeLowercase{\textit{et al.}}: A Sample Article Using IEEEtran.cls for IEEE Journals}

\maketitle

% due to the spatio-temporal interaction of mixed traffic environment, it is still difficult for the existing decision-making systems of CAVs to make accurate judgments and effective strategies.
\begin{abstract}
In the realm of heterogeneous mixed autonomy, vehicles experience dynamic spatial correlations and nonlinear temporal interactions in a complex, non-Euclidean space. These complexities pose significant challenges to traditional decision-making frameworks. Addressing this, we propose a  hierarchical reinforcement learning framework integrated with multilevel graph representations, which effectively comprehends and models the spatiotemporal interactions among vehicles navigating through uncertain traffic conditions with varying decision-making systems. Rooted in multilevel graph representation theory, our approach encapsulates spatiotemporal relationships inherent in non-Euclidean spaces. A weighted graph represents spatiotemporal features between nodes, addressing the degree imbalance  inherent in dynamic graphs. We integrate asynchronous parallel hierarchical reinforcement learning with a multilevel graph representation and a multi-head attention mechanism, which enables connected autonomous vehicles (CAVs) to exhibit capabilities akin to human cognition, facilitating consistent decision-making across various critical dimensions. The proposed decision-making strategy is validated in challenging environments characterized by high density, randomness, and dynamism on highway roads. We assess the performance of our framework through ablation studies, comparative analyses, and spatiotemporal trajectory evaluations. This study presents a quantitative analysis of decision-making mechanisms mirroring human cognitive functions in the realm of heterogeneous mixed autonomy, promoting the development of multi-dimensional decision-making strategies and a sophisticated distribution of attentional resources. 
% By combining parallel asynchronous hierarchical reinforcement learning with the framework, we equip Connected and Autonomous Vehicles (CAVs) with the ability to perform cognitive multitasking similar to human cognitive processes, enabling them to handle multiple decision-making tasks concurrently across different dimensions. Experimental validation in a complex, unpredictable, and dynamic traffic environment shows that our framework, enhanced with graph attention mechanisms, effectively replicates human-like distributed attention and multitasking abilities. 
% This research offers quantitative insights into human-like decision-making processes within heterogeneous mixed autonomy contexts, facilitating parallel learning of multi-dimensional decision strategies with limited attention resources.
\end{abstract}

\begin{IEEEkeywords}
Heterogeneous mixed autonomy, connected autonomous vehicles,  Multi-level Graph theory, parallel asynchronous hierarchical reinforcement learning. 
\end{IEEEkeywords}

\IEEEpeerreviewmaketitle

\makenomenclature
\mbox{}
\nomenclature{ \( N \)}{Total number of vehicles in   traffic environment}
\nomenclature{\( M \)}{Total number of CAVs}
\nomenclature{ \( \mathcal{T} \)}{Dimension of  graph}
\nomenclature{\(L\)}{Lane-changing dimension}
\nomenclature{ \(F\) }{Following dimension}
\nomenclature{\(\mathcal{G}_{\mathcal{T}}\)}{Graph representation for traffic environment}
\nomenclature{ \( V \)  }{Set of nodes in  graph}
\nomenclature{ \( E \)}{Set of edges in  graph}
\nomenclature{\( N_{ \mathcal{T} -t}\) }{Sub-node feature matrices for \( \mathcal{T} \) dimension}
\nomenclature{ \( A_{\mathcal{T} - t} \)}{Difference in time-steps between  decision-making dimensions}
\nomenclature{ \(\Delta\)}{Sub-adjacency matrices matrixs for \( \mathcal{T} \) dimension}
\nomenclature{ \(\mathcal{S}\)}{State information in current traffic environment}
\nomenclature{ \(\mathcal{A}_{\mathcal{T}}\)}{All possible joint actions \((a_L, a_F) \)}
\nomenclature{ \(\mathcal{P}\)}{Transition probability}
\nomenclature{ \(\mathcal{R}_{\mathcal{T}}\)}{Sum of all rewards from CAVs for \( \mathcal{T} \) dimension}
\nomenclature{ \(\kappa_{ij} \)}{Standard safety distance coefficient between vehicles \(i\) and   \(j\)}
\nomenclature{ \( e_{ij}^k \)}{Normalized attention coefficient}

\printnomenclature

\section{Introduction}

% This file is intended to serve as a template for IEEE Transactions on Vehicular Technology manuscripts under \LaTeX\ IEEEtran.cls version 1.8a and later.

% \hfill August 4, 2015

\IEEEPARstart{A}{utomobile} automation  is emerging as a pivotal trend  \cite{Baruch}. In the United States, 62 companies specializing in autonomous driving have been authorized to perform road tests in California. It is anticipated that by 2040, 40\% of China's   vehicles will achieve automation. Road traffic systems are expected to remain in a state of transition for the foreseeable future, with the simultaneous presence of vehicles with diverse levels of intelligence, including Human-driven Vehicles (HVs), Connected Human-driven Vehicles (CHVs), and connected and autonomous vehicles (CAVs) \cite{9310545}. Such diversity introduces  significant  heterogeneity in terms of intelligence, communication capabilities, decision-making processes, and driving tasks \cite{8744265}, posing   challenges to   collaborative micro-level decision-making.

Efficient navigation in diverse and intricate traffic scenarios is essential for the advancement of autonomous driving technologies. These scenarios are defined by heavy traffic, the convergence of vehicles and pedestrians, unpredictable behaviors, and numerous operational restrictions. We term the operation of autonomous vehicles in such settings as heterogeneous mixed autonomy. Present decision-making frameworks based on expert rules exhibit a lack of flexibility, and struggle to adjust to varying scenarios. In the sphere of heterogeneous mixed autonomy, the intelligence of autonomous vehicles does not yet match that of human drivers, which compromises traffic flow and jeopardizes road safety. This area is distinguished by dynamic spatial correlations and nonlinear temporal correlations. Dynamic spatial correlation highlights the changing relationship of traffic patterns across   locations over time, and nonlinear temporal correlation reflects a nonlinear relationship between traffic states and their timing. These nonlinear spatiotemporal interactions among entities occur in non-Euclidean spaces. Hence, conventional deep learning approaches, such as Convolutional Neural Networks (CNNs) and Long Short-Term Memory  (LSTM) networks, fail to adequately capture these intricate spatiotemporal relationships.

CAVs embody advanced systems characterized by substantial self-learning and iterative improvement capabilities.  In a nascent stage of intelligence, they are primarily suited to structured road scenarios with sparse traffic flows. In urban traffic conditions, CAVs still must effectively integrate spatiotemporal dynamics with nearby vehicles and develop self-learning abilities to navigate unknown scenarios. Within the framework of heterogeneous mixed autonomy, CAVs demonstrate elementary cognitive functions comparable to humans, and  significant dependence on human intervention. Insights from cognitive psychology and allied fields suggest that the skillfulness of biological intelligence at solving problems stems from hierarchical cognitive processes \cite{BOTVINICK2009262}, and  it becomes evident that a similar approach could be beneficial in the domain of autonomous vehicles \cite{10250993}. Hierarchical reinforcement learning is a promising computational approach, potentially equipping CAVs with problem-solving skills mirroring those of biological intelligence  \cite{eppe2022intelligent}, effectively enabling the segmentation of decision-making tasks into separate dimensions.  Yet, in   complex and dense traffic scenarios, human drivers still markedly outperform artificial systems. The key factors contributing to this gap include: (a)  inadequately   capturing the full spectrum of vehicle interactions within diverse dimensions; and (b) a deficiency in consistent cognitive processing for decision-making across these dimensions. This leads to the question of how   CAVs can be equipped with a hierarchical learning capability that parallels the consistent cognitive abilities observed in humans.

We propose a hierarchical reinforcement learning framework based on multilevel graph representation theory, to accurately capture the spatiotemporal dynamics in the realm of heterogeneous mixed autonomy. This facilitates decision-making processes reflecting human cognitive strategies by  managing the complexities of spatiotemporal relationships, particularly in non-Euclidean spaces. A  parallel asynchronous hierarchical graph reinforcement learning strategy is designed to   enhance the self-learning capabilities and iterative improvement processes of CAVs. We employ specialized node feature matrices, dynamic adjacency matrices, and   reward functions to   analyze various critical dimensions. This   ensures an optimized distribution of attentional resources, mirroring the human capacity for divided attention and adept multitasking. We assess the viability and effectiveness of our proposed decision-making framework in a simulated environment of high-density, unpredictable, and dynamic highway   traffic. The principal contributions of this paper are outlined as follows:
\begin{enumerate}[(a)]
\item A multilevel graph representation theory is proposed to model the spatiotemporal dependencies present in non-Euclidean spaces;
\item Weighted graphs based on spatiotemporal features between nodes address the degree imbalance problem in dynamic graphs. The coupling of multidimensional graphs with multi-head attention mechanisms enables human-like divided attention and consistent cognitive capabilities;
\item A parallel asynchronous hierarchical graph reinforcement learning approach is proposed to enhance the self-learning iteration capabilities of CAVs. This includes an asynchronous multidimensional Markov decision process and  improved GRL algorithms based on multilevel graphs;
\item  The comprehensive performance of the proposed framework is tested through ablation studies, comparative experiments, and spatiotemporal trajectory analysis, which show  that CAVs are capable of inter-dimensional collaboration, and hence to make cognitive decisions.
\end{enumerate}

% Aiming at the shortcomings of the existing works,  a discrete decision strategy is proposed for autonomous driving based on a rate graph convolution Q-learning network (Rate GQN) to improve the comprehensive performance and stability of driving. The main contributions of this paper are as follows:

% \subsubsection{Improved RL algorithm with mathematical proofs} An improved algorithm based on DQN, Rate DQN, is innovatively proposed. Specifically, the estimates of the Q-values in the previous learning process is ratioed, thus improving the stability and performance of the algorithm by reducing the approximation error of the target values. Furthermore, mathematical proofs are given to show that the method can reduce the variance of the target approximation error.

% \subsubsection{Interaction represented by graph convolutional network}The graph convolutional network  extracts features in the traffic scenario, including information exchanged between different vehicles and information transmitted to vehicles by signal lights. These features help CAVs  grasp the global traffic state and local interaction information quickly and accurately.

% \subsubsection{An internal dynamic multi-objective reward function}
% An internal dynamic multi-objective reward function is designed to improve the overall performance of CAVs, including safety, efficiency, energy saving, and comfort. Simulation results validate the effectiveness and robustness of our solution in terms of learning efficiency, decision accuracy, and driving performance.

The remainder of this article is organized as follows. Section \RNum{2} introduces   related work. The proposed multilevel graph representation theory is illustrated in section \RNum{3}. Section \RNum{4} introduces the methodology, including an asynchronous multidimensional graphical Markov decision process and parallel asynchronous hierarchical reinforcement learning. Section \RNum{5} explains the   setting of   experiments. Section \RNum{6}   analyzes the simulation results. Conclusions are drawn in section \RNum{7}.

\section{Related Work}

% \begin{figure*}[bt!]
%     \centering
%     \includegraphics[width=\linewidth]{schematic.eps}
%     \caption{ The schematic diagram of the designed framework. Firstly, the scenario is characterized by a graph representation. In this representation, all vehicles are considered as nodes, and their local information, as well as global information such as the status of traffic lights on the road they are traveling, are treated as node attributes. The communication between vehicles is modeled as edges in the graph.Subsequently, the graph matrices are processed through graph neural networks and deep neural networks to extract features, which are then used to determine lane changing and speed adjustment actions for CAVs. At each time step, the environment's state information changes, and the reward value is calculated based on the updated state information. Notably, the energy module calculates the energy consumption of CAVs, which is taken into consideration when calculating the reward function.Finally, the proposed Rate algorithm is employed to continuously train the neural network model offline, aiming to optimize the decision-making strategy.}
%     \label{fig1}
% \end{figure*}

\subsection{Graph Representation for Connected and Autonomous Vehicles}
% Scheme description goes here \cite{IEEEhowto:kopka}.
Scholars have identified limitations in traditional decision-making methodologies based on Multilayer Perceptron (MLP) and LSTM that primarily focus on individual vehicle characteristics, neglecting   interactive dynamics among neighboring vehicles \cite{8917228}. To address this, several researchers have adopted the Graph Neural Network (GNN) to represent dynamic traffic scenarios \cite{peng2022drl,dong2020drl}, where   vehicles are depicted as nodes in a graph, with   inter-vehicular relationships portrayed as edges between   nodes. Hart et al. \cite{9304738} employed a GNN to display these vehicular interactions, and found that a GNN, in contrast to conventional deep neural networks, can well handle varying numbers of vehicles across diverse scenarios, significantly enhancing the versatility of decision-making models.  Chen et al. \cite{Chen2021838} conceptualized the interaction of CAVs in traffic scenarios as a dual-layer network, where the first layer is a star-shaped local network, encompassing CAVs and their adjacent HVs, and the second layer is a global network, consisting of nodes that correspond to CAVs on roadways. By integrating a Graph Convolutional Network (GCN) and Deep Q-Network (DQN) in a combined GCQ model, they facilitated collaborative lane-changing decisions among CAVs.

Several   studies have further explored graph representations of interactions among CAVs. Hu et al. \cite{10164220} introduced a GCN featuring a trainable adjacency matrix,   addressing the   ineffective representation   of dynamically changing relationships between nodes. Hu et al. \cite{8638814}  used  dynamic coordination graphs to model the evolving topological structures inherent in vehicle interactions, and two fundamental learning methods to coordinate the driving maneuvers of a group of vehicles, thus  enhancing the coordination of decision-making.   Gao et al. \cite{Gao20231} included additional dimensions of vehicular interactions,  representing interactions between vehicles, and  those between traffic signals and CAVs, as edges and nodes, respectively. Experimental results indicated that the method enables CAVs to rapidly and accurately grasp   global traffic states and local interaction data.

In the realm of CAVs, while previous research on graph representations has adeptly modeled the interactions and relationships among multiple vehicles, these studies have not delved deeply into the spatiotemporal interactive behaviors of CAVs. Research has primarily provided a superficial representation of spatial interactions, focusing only on the immediate neighbors of a target vehicle and neglecting the dynamic spatial correlations inherent in heterogeneous mixed autonomy. Recent work has introduced dynamic graphs to represent the temporal correlations of CAVs. However,  vehicles that suddenly exceed a predefined distance threshold   can cause abrupt   topological changes, resulting in the disappearance of nodes. This type of step-like interaction essentially embodies a nonlinear temporal correlation, which can affect robustness in decision-making processes.

\subsection{Decision-making based on Hierarchical Reinforcement learning}
Deep Reinforcement Learning (DRL) offers a paradigm shift in learning complex decision-making strategies, as it does not rely on a large amount of labeled driving data \cite{mnih2015human,wu2023human,he2023fear},   learning instead through direct interaction with the environment. This approach has been extensively applied in traffic scenarios to address the autonomous decision-making challenges faced by CAVs. Various methods employing DRL have been developed to train CAVs for highway merging tasks, aiming to enhance traffic throughput while maintaining zero collisions \cite{9575676,10159552}. Other techniques focus on training CAVs for safe and efficient decision-making at urban intersections \cite{9304606,8569400,9700479}. Notable among these methods are   risk assessment models \cite{8813791,LI2022103452}, combinatorial decision strategies \cite{9712209,9700479}, and multitask reward models \cite{9304542,9497870}. Much   research concentrates on the performance of autonomous vehicles in singular, controlled scenarios, but real-world traffic environments present   far more dynamic and complex scenarios.

Hierarchical Reinforcement Learning (HRL) can well manage complex, dynamic traffic scenarios by introducing a hierarchical structure in the decision-making process \cite{bacon2017option}. HRL decomposes the decision-making of CAVs into distinct hierarchical layers, simplifying the learning of complex tasks, and enabling CAVs to more efficiently solve large-scale and intricate problems. A hierarchical DRL framework was   constructed by dividing driving tasks into sub-tasks,   allocated to different sub-models for training, which are subsequently integrated. Experiments   demonstrated a 100\% success rate in task completion  \cite{9497870}. Duan et al. \cite{https://doi.org/10.1049/iet-its.2019.0317} proposed an HRL method that does not depend on a large volume of labeled driving data. The driving task is dissected into   strategies of driving in lane, right lane change, and left lane change. A master strategy is   learned, so as to select the appropriate maneuvering strategy in the current state.  Peng et al. \cite{9843863} introduced   upper- and lower-level deep reinforcement learning algorithms to address complex state spaces, employing D3QN and DDPG to train lane-changing and following decisions, respectively. The   model was found to increase the driving speed by 23.99\%.

Although previous HRL methods have successfully achieved the decomposition of decision-making tasks into distinct dimensions,   deeper exploration is needed across various dimensions, and a re-examination and study of vehicle interaction relationships within these   decision dimensions are imperative. For instance, in the lane-changing and tracking dimensions, emphasis should be placed on interactions with vehicles in adjacent lanes and  in the same lane, respectively.  HRL decision-making methodologies have begun to adopt parallel learning strategies. But the synchronous parallel learning approach disregards the cognitive integration across diverse decision-making dimensions. For instance, within environments characterized by heterogeneous mixed autonomy, CAVs   must adeptly modulate their acceleration to identify a secure spatiotemporal region before changing  lanes.
\section{Multilevel Graph Representation Theory}
We introduce the proposed multi-level graph representation theory. The foundational theory of traffic graphs   includes the mathematical representation of the node feature matrix and adjacency matrix.  Expanding upon the principles of divided attention and  cognitive consistence, we explore traffic graph theory, broadening its applicability to dynamic, weighted, and multidimensional graph theories \cite{10077454}.

\subsection{Foundational Theory of Traffic Graphs}\label{section:3.1}

Assume a scenario where, at each discrete time-step \( t \), the traffic environment is comprised of \( N \) vehicles, where \( M \) (\( M \leq N \)) of those are autonomous vehicles that interact with others. The traffic scenario  is conceptualized as a graph \( G_t(V, E) \) with \( N \) nodes, where \( V(G) \) denotes the collection of nodes within the graph,  each node \( u \) (\( u \in V \)) representing a vehicle; \( E(G) \) represents the set of edges, illustrating the exchange of information between vehicles as edges \( (u, v) \in E \) \cite{Gao20231,liu2022graph}. The   graph at time-step \( t \) is characterized by   the node feature matrix \( N_t \in \mathbb{R}^{N \times F} \) and   adjacency matrix \( A_t \in \mathbb{R}^{N \times N} \), where \( F \) is the total number of features attributed to each node.

\subsubsection{Node Feature Matrix}
The node feature matrix \( N_t \) indicates the states of all vehicles, including   CAVs  and HVs. For the ${i_{th}}$ vehicle, the state matrix \( S_t \) indicates   parameters such as vehicle speed \( V_i \), lateral position \( X_i \), longitudinal position \( Y_i \), road segment occupancy \( R_i \), lane occupancy \( L_i \), and vehicle category \( I_i \). The mathematical representation is
\begin{equation}
\left\{
\begin{array}{ll}
S_t = \left[ V_i, X_i, Y_i, R_i, L_i, \cdots, I_i \right] \\
N_t = \left[ S_t^1, S_t^2, \cdots, S_t^i, \cdots, S_t^N \right]
\end{array}
\right.,
\end{equation}
where \( S_t^i \) (\( i = 1,2, \cdots, N \)) is the feature set of the ${i_{th}}$ vehicle at the ${t_{th}}$ time-step.

\subsubsection{Adjacency Matrix}
The interactions between vehicles at the ${t_{th}}$ time-step are represented by the adjacency matrix \( A_t \in \mathbb{R}^{N \times N} \), 
\begin{equation}
{A_t} = \left[ {\begin{array}{*{20}{c}}
{{a_{11}}}&{{a_{12}}}& \cdots &{}& \cdots &{{a_{1N}}}\\
{{a_{21}}}&{{a_{22}}}& \cdots &{}& \cdots &{{a_{2N}}}\\
 \vdots & \vdots & \ddots &{}&{}& \vdots \\
{}&{}&{}&{{a_{ij}}}&{}&{}\\
 \vdots & \vdots &{}&{}& \ddots & \vdots \\
{{a_{N1}}}&{{a_{N2}}}& \cdots &{}& \cdots &{{a_{NN}}}
\end{array}} \right],
\end{equation}
a binary matrix, where \( a_{ij} \) has a value of 1 or 0 to indicate the presence or absence, respectively, of a connection between the \(i\)th and \(j\)th nodes. 

Based on the principles of directed graphs, the computation of the adjacency matrix is based on four assumptions: a) CAVs  within a defined distance threshold on adjacent lanes can share information; b) Information cannot be shared between HVs; c) All CAVs can share information about HVs within their threshold area; d) A vehicle can share information with itself, i.e., \( a_{ii} = 1 \).

\begin{figure*}[bt!]
    \centering
    
    \subfigure[Local schematic representation of a dynamic traffic graph.]{
        \begin{minipage}[t]{0.53\linewidth}
            \includegraphics[width=9cm]{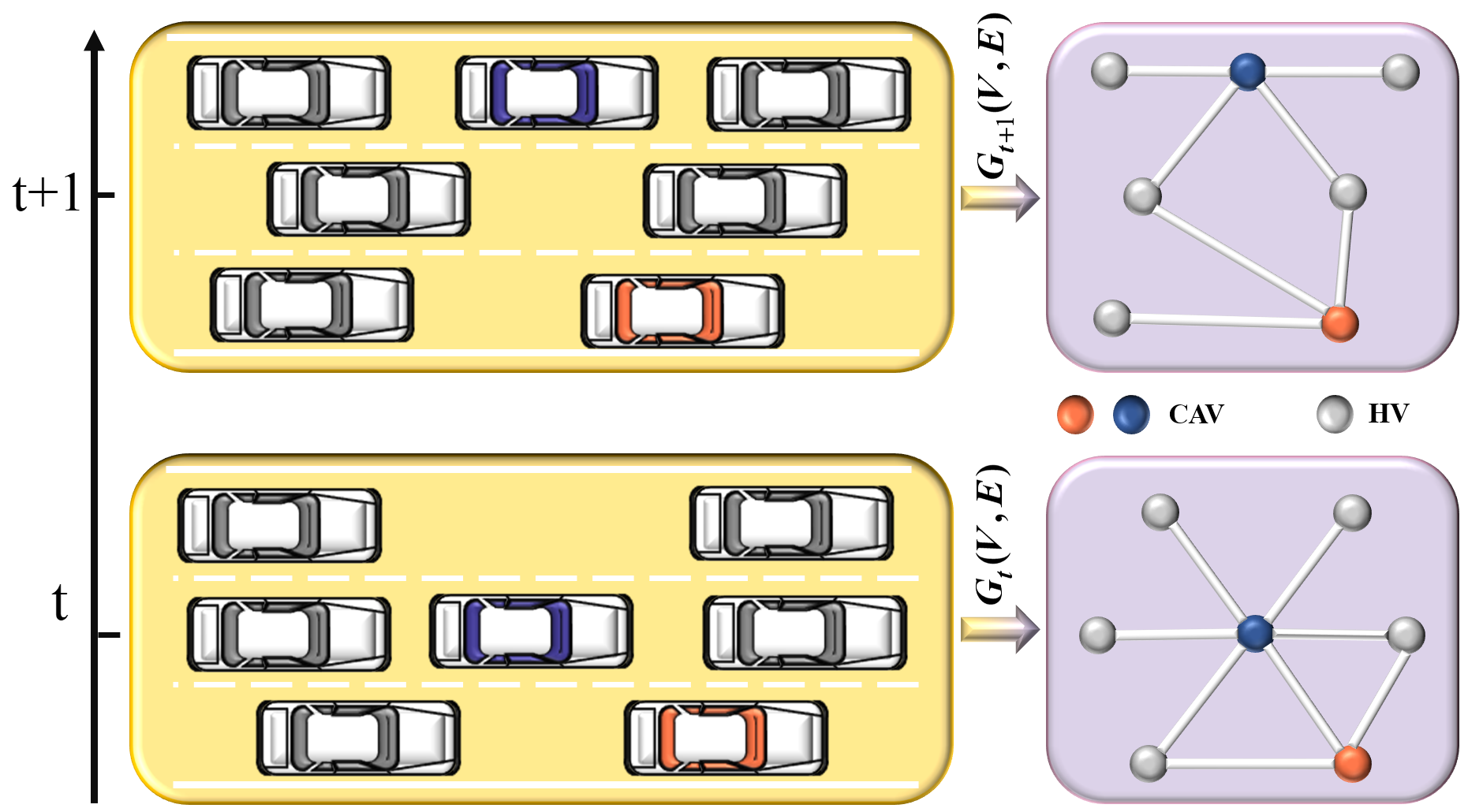}
            
        \end{minipage}
    } 
    \subfigure[Local schematic representation of a multidimensional traffic graph.]{
        \begin{minipage}[t]{0.435\linewidth}
            \includegraphics[width=7cm]{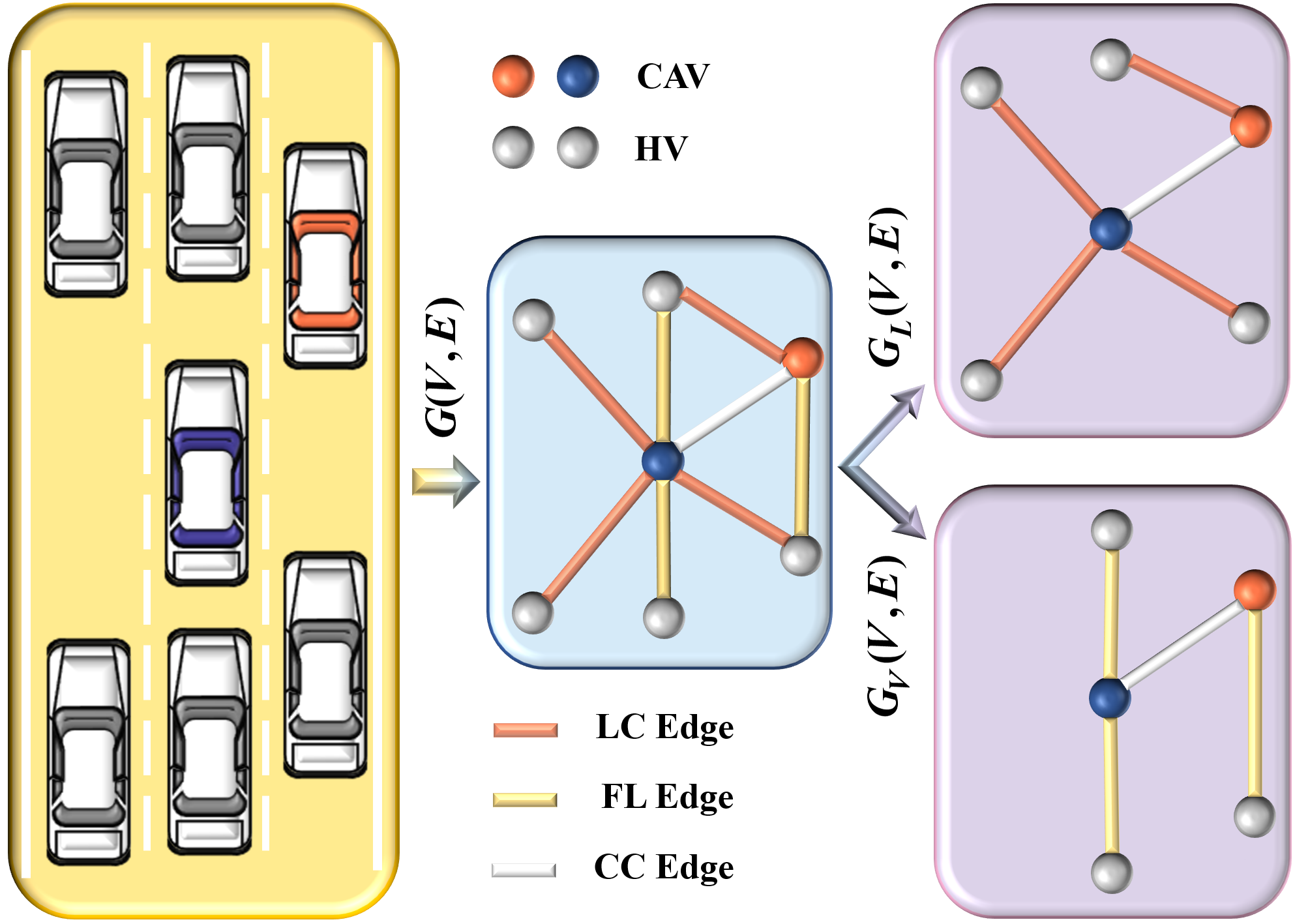}
            
        \end{minipage}
    }
    
    \caption{Schematic diagrams of  dynamic   and  multidimensional traffic graphs. Colored and gray circles denote CAVs and HVs, respectively. (a)  Spatial dynamics among vehicles alter   graph's topological structure from \(t\) to \(t + 1\), affecting number of nodes and edges, and objectives of edge connections, while categorization of edges remains unchanged; (b)   Interactions are distinguished based on   vehicle type and spatial dynamics. LC: interaction between  CAV and   HV in   adjacent lane; FL:  interaction between  CAV and   HV in same lane; CC:  interaction between  two CAVs in adjacent lanes or  same lane.}
    \label{fig:2}
\end{figure*}
\subsection{Multilevel Graph Representation Theory}
In the context of heterogeneous mixed autonomy, dynamic spatial correlations and nonlinear temporal interactions among diverse vehicular entities are observed within a non-Euclidean domain, in which the multilevel structure of the graph can well capture dependencies. Graphs constituting the traffic scenario can be categorized as directed or undirected, static or dynamic, unidimensional or multidimensional, and uniform or weighted. We discussed directed graphs above, and below we address  dynamic, multidimensional, and weighted graphs in traffic scenarios.

% \begin{figure}[!t]
% \centering
% \includegraphics[width=3.25in]{figure2.png}
% \caption{ The dynamic traffic graph.}
% \label{fig.2}
% \end{figure} 

\subsubsection{ Dynamic Traffic Graph}

In traffic scenarios, the mobility of vehicles results in changes to both the individual vehicle characteristics and the topological structure of the graph, rendering it  dynamic. In the dynamic traffic graph, based on changes in its spatiotemporal properties, we identify two dynamic characteristics: a) New vehicle nodes can appear or disappear, and new interactions between vehicles can be established or terminated within different time periods; b) The movement of vehicles leads to changes in   their individual characteristics  and the nature of their interactions.

\begin{definition}[Dynamic Traffic Graph]

A dynamic traffic graph is defined as \( G = (G_1, G_2, \cdots, G_t, \cdots, G_T) \), and at time-step \( t \) is defined as \( G_t = G_t(V_t, E_t) \), where \( t \in T \), and \( T \) is the total number of time-steps. Each edge \( e_{ij}^t = E_t \) at time-step \( t \) connects the \( i \)-th and \( j \)-th vehicles in interaction. Each node is represented by the state information of a vehicle at the current moment \( t \). Figure \ref{fig:2}(a) shows a local schematic representation of a dynamic traffic graph.

\end{definition}
When a node disappears in a dynamic graph, its edges with other nodes are severed, directly impacting the graph's structural characteristics, especially its degree distribution. In uniform graphs, whose adjacency matrix values are only 0 or 1, with the disappearance of connections between nodes,   related edges transition from 1 to 0, which is referred to as a step-wise imbalance in the graph's degree distribution. Appendix A shows  the   mathematical analysis.

Theorem 1 addresses this issue, and is proved in Appendix B.
\begin{theorem}[Equilibrium Theorem for Degree Distribution in Weighted Dynamic Graphs]
Upon transitioning from time \(t\) to \(t+1\), within a dynamic graph \(G_t\) characterized by a weighted adjacency matrix \(A_t\)--with weights \(a_{ij}^t\) stemming from both spatial distances and attributes of nodes--the disruption in degree distribution prompted by the elimination of a node \(u\) is alleviated by the graph's weighted mechanism. In particular, the alteration in the weighted degree of any node \(v\) resultant from node \(u\)'s elimination undergoes a seamless adjustment, conforming to the formula
\begin{equation}
d_{t+1}^{w}(v) = d_t^{w}(v) - a_{uv}^t,
\end{equation}
which facilitates a steadier degree distribution transition than that typically encountered in unweighted (binary) graphs.
\end{theorem}

% \begin{theorem}
% During the transition from time \(t\) to \(t+1\), if a node \(u\) is removed from the graph \(G_t\), any resultant imbalance in the degree distribution can be alleviated or rectified by employing a weighted adjacency matrix.
% \end{theorem}
% \begin{figure}[!t]
% \centering
% \includegraphics[width=2.5in]{figure3.png}
% \caption{ The multi-dimensional traffic graph.}
% \label{fig.3}
% \end{figure} 
\subsubsection{ Multidimensional Traffic Graph}\label{section:3.2}

In dynamic traffic scenarios, the interactions between vehicles exhibit dynamic spatial correlations. As vehicles move through space, their modes of interaction change, and can have the following three characteristics: a) Interactions between CAVs  and HVs  on adjacent lanes are considered lane-changing (LC) edges; b) Interactions between CAVs and HVs on the same lane are considered as following (FL) edges; and c) Interactions between CAVs on adjacent lanes are considered connected CAV-CAV (CC) edges. The CC edges form the basis of a multidimensional traffic graph through expansion to LC and FL edges.
\begin{definition}[Multidimensional Traffic Graph]
Define the traffic environment at each time \( t \) as a multidimensional graph \( G_t(\mathcal{T}, V, E) \), where \( \mathcal{T} \in (L, F) \) represents the dimensions of the graph, \( V(G) \) is the set of nodes, and \( E(G) \) is the set of edges. Depending on the graph dimension \( \mathcal{T} \),  graph \( G(\mathcal{T}, V, E) \) is decoupled into a lane-changing dimension graph \( G_L(V_L, E_L) \) and a following dimension graph \( G_F(V_F, E_F) \). Subsequently, for temporal representation of the graph, at time-step \( t \), it is characterized by the node feature matrix \( N_{\mathcal{T} - t}^{M \times F_{\mathcal{T}}} \) and   adjacency matrix \( A_{\mathcal{T} - t}^{M \times M} \), where \( F \) is the total number of features per vehicle, and \( M \) is the total number of vehicles. The sub-node feature matrices for different dimensions \( \{ N_{L - t}^{M \times F_L}, N_{F - t}^{M \times F_F} \} \in N_{\mathcal{T} - t}^{M \times F} \) are extracted, where \( F_{\mathcal{T}} \le F \). Based on   vehicle interactions in different dimensions \( (u_{\mathcal{T}}, v_{\mathcal{T}}) \in E_{\mathcal{T}} \),   adjacency matrices are similarly decoupled into two dimensions \( \{ A_{L - t}^{M \times M}, A_{F - t}^{M \times M} \} \). Figure \ref{fig:2}(b) shows a local schematic representation of a multidimensional traffic graph.
\end{definition}

The proposed multidimensional traffic graph accounts for  lane-changing and following, which enables precise capture of   spatiotemporal interactions between vehicles across different dimensions. Coupled with the multilevel multi-head attention mechanism (section \ref{section4.3}), it is possible to emulate human-like divided attention and consistent cognitive  capabilities, such as observing the surrounding environment and preparing for lane changes while maintaining lane discipline.This nuanced approach to modeling vehicular dynamics serves as a precursor to the Theorem 2, which demonstrates the systematic reduction in entropy within sub-dimensional graphs, underscoring the efficacy of targeted attention in complex traffic scenarios.

\begin{theorem}[Targeted Attention-Driven Entropy Reduction in Sub-Dimensional Graphs]
Consider a complex dynamic graph \(G\) that encompasses multi-dimensional representation, along with its corresponding sub-dimensional graphs \(G_{\mathcal{T}}\), wherein \(\mathcal{T} \in \{G_L, G_F\}\) signifies distinct dimensions of interaction. For any nodes \(i\) that is congruent with decision-making objectives in \(G_{\mathcal{T}}\), it is posited that the probability of attention \(p_{\mathcal{T}}(i)\) surpasses the analogous probability across the global graph \(G\), represented by \(p(i)\), Formally, 
\begin{equation}
p_{\mathcal{T}}(i) > p(i), \quad \forall i \in V_{\mathcal{T}} 
\end{equation}   

Based on this premise, the entropy related to edge attention in each dimension-specific subgraph, \(H_{\mathcal{T}}\), is demonstrably lower than the entropy found in the global graph, \(H_{G}\). This relationship is defined mathematically as
% \begin{equation}
% H_{\mathcal{T}} = - \sum_{i \in V_{\mathcal{T}}} p_{\mathcal{T}}(i) \log p_{\mathcal{T}}(i) < H_{G} = - \sum_{i \in V} p(i) \log p(i).
% \end{equation} 
% \begin{equation}
% H_{\mathcal{T}}=-\sum_{i\in V_{\mathcal{T}}}p_{\mathcal{T}}(i)\log p_{\mathcal{T}}(i),
% H_G=-\sum_{i\in V}p(i)\log p(i),
% H_{\mathcal{T}}<H_{G}.
% \end{equation}
\begin{align}
H_{\mathcal{T}} &= -\sum_{i\in V_{\mathcal{T}}} p_{\mathcal{T}}(i)\log p_{\mathcal{T}}(i), \\
H_G &= -\sum_{i\in V} p(i)\log p(i), \\
H_{\mathcal{T}} &< H_{G}.
\end{align}

\end{theorem} 

This conclusion substantiates that the focused allocation of attention towards nodes that are directly relevant to the decision-making objectives in each sub-dimension results in a significant reduction in entropy, \(H_{\mathcal{T}}\), evidencing a more concentrated and efficient distribution of attention within sub-dimensional graphs compared to the comprehensive graph.

\subsubsection{Weighted Traffic Graph}

In traffic scenarios, a weighted graph can represent spatiotemporal interactions between vehicles with greater precision than a uniform graph.  It can more finely characterize the dynamic spatial interactions between vehicles, as the edge weights can represent not only the existence of a connection between vehicles, but can include multidimensional information such as distance, relative speed, or traffic density. Also, weighted graphs can more effectively simulate nonlinear temporal correlations. In uniform graphs, the relationships between vehicles are often simplified as binary. In weighted graphs, the edge weights change with the states of   vehicles, and hence can represent the strength and timeliness of vehicle interactions. Based on the proposed dynamic and multidimensional traffic graph, weight representations are determined separately   for the two-dimensional dynamic graphs. The adjacency matrices of the lane-changing dimension graph and   following dimension graph \(\{ A_{WL - t}^{M \times M}, A_{WF - t}^{M \times M} \}\) are individually designed.

The sub-adjacency matrix \( A_{WL - t}^{M \times M} \) for the lane-changing dimension graph is constructed based on the distance between vehicles and their lanes. Considering the difficulty human drivers face in estimating the speed of vehicles in adjacent lanes, the weight \( a_{ij}^L \) is defined as a function of the distance and lane,  decreasing with  distance, and ranging between 0 and 1. For the ${j_{th}}$ vehicle located on an adjacent lane to the ${i_{th}}$ CAV, with a lane-direction distance of \( \Delta d_{ij} \) and a distance threshold \( X \), the weight function is defined as
\begin{equation}\label{weighted_L}
a_{ij}^L = \left\{
\begin{array}{ll}
\max(0, 1 - \frac{\Delta d_{ij}}{X}), & \text{if } d_{ij} < X, \\
0, & \text{if } d_{ij} \ge X.
\end{array}
\right.
\end{equation}

As the distance \( \Delta d_{ij} \) approaches the threshold \( X \), the weight approaches 0. When   \( \Delta d_{ij} \) is small, indicating close proximity, the weight approaches 1.

For the sub-adjacency matrix \( A_{WF - t}^{M \times M} \) of the following dimension graph, which is built based on the distance and speed difference between two vehicles on the same lane, the weight \( a_{ij}^F \) is  a function of both the distance and speed difference, decreasing with an increase in distance, and a decreased   speed difference, and ranging between 0 and 1. For the \( j \)-th vehicle   on the same lane as the \( i \)-th CAV, with a lane-direction distance of \( \Delta d_{ij} \) and  distance threshold \( Y \), the weight function is defined as
\begin{equation}\label{weighted_F}
a_{ij}^F = \left\{ 
\begin{array}{ll}
\exp\left( - \frac{\Delta d_{ij}}{\lambda_d} - \frac{\lambda_v}{\Delta v_{ij}} \right), & \text{if } d_{ij} < Y \text{ or } \Delta v_{ij} \geq \Delta V, \\
0, & \text{else}.
\end{array} 
\right.
\end{equation}

As \( \Delta d_{ij} \) approaches the distance threshold \( X \), and \( \Delta v_{ij} \) approaches the speed threshold \( \Delta V \), the weight approaches 0, and conversely, the weight approaches 1 as \( \Delta d_{ij} \) decreases and \( \Delta v_{ij} \) increases.
\begin{figure*}[bt!]
    \centering
    \includegraphics[width=0.85\linewidth]{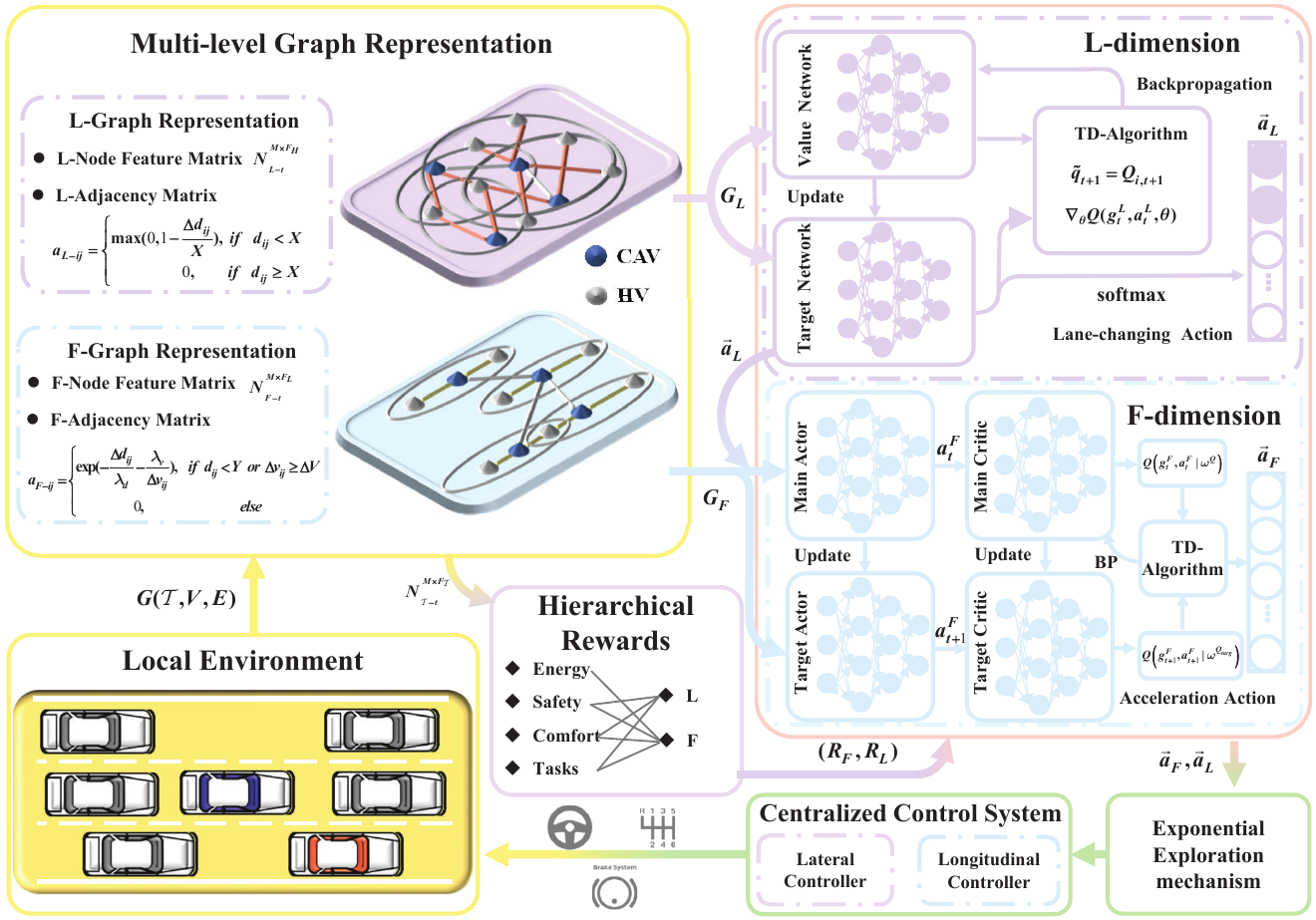}
    \caption{Proposed multilevel graph reinforcement learning framework. Traffic scenario is represented as  multilevel graph. Based on   characterization   of its multidimensional graphs,  parallel asynchronous hierarchical graph reinforcement learning algorithm is proposed for training actions in both   lane-changing and following dimensions. Actions are enhanced in   exploration efficiency through   exponential exploration mechanism, and subject to centralized control system. Hierarchical reward functions computes reward feedback for corresponding dimensions, facilitating policy iteration updates.}
    \label{fig:frame}
\end{figure*}
\section{METHODOLOGY}
We discuss the methodology employed in this study. The problem formulation includes an asynchronous multidimensional graphical Markov decision process (AMG-MDP), with   state spaces, action spaces, and reward functions across different dimensions. The proposed Parallel Asynchronous Hierarchical Reinforcement Learning (PAHRL) strategy and multilevel multi-head graph attention network module  are described. Multilevel graph reinforcement learning is illustrated in Figure \ref{fig:frame}.
\subsection{Problem Formulation}
\subsubsection{AMG-MDP}
\begin{figure}[!t]
\centering
\includegraphics[width=3.5in]{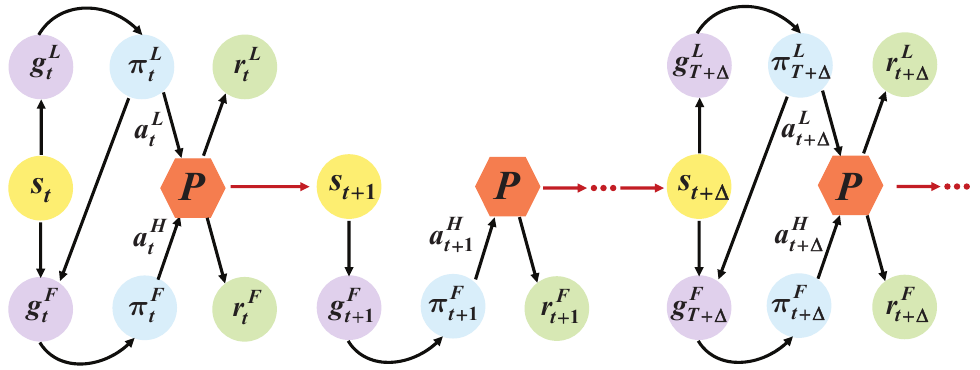}
\caption{Asynchronous Multidimensional Graphical Markov Decision Process.}
\label{fig.3}
\end{figure} 
Heterogeneous mixed autonomy presents novel challenges for the coordination of cross-dimensional decisions by CAVs.   AMG-MDP modifies the conventional MDP framework  to model the spatiotemporal correlations in CAV decisions across various dimensions,  allowing the MDP framework to accommodate multi-agent coupled decision-making.

\begin{definition}[AMG-MDP]
The Asynchronous Multidimensional Graphical Markov Decision Process  can be defined by the tuple \(\mathcal{N} = (n, m, \mathcal{T}, \Delta, \mathcal{S}, \mathcal{G}_{\mathcal{T}}, \mathcal{A}_{\mathcal{T}}, \mathcal{P}, \mathcal{R}_{\mathcal{T}})\), where \(n\) is the number of vehicles present in the current scenario \(\{V_1, V_2, \ldots, V_n\}\); \(m\) is the number of CAVs \(\{CAV_1, CAV_2, \ldots, CAV_m\}\); \(\mathcal{T} \in (L, F)\) gives the decision dimensions of   AMG-MDP, where \(L\)      and \(F\) are respectively the   lane-changing and following dimensions; \(\Delta\) is the difference in time-steps between decisions in the two dimensions; \(\mathcal{S}\) is the state information of all vehicles in the current environment, where \((s_1^{\mathcal{T}}, s_2^{\mathcal{T}}, \ldots, s_i^{\mathcal{T}}, \ldots, s_n^{\mathcal{T}}) \in \mathcal{S}_{\mathcal{T}}\); \(\mathcal{G}_{\mathcal{T}}\) is the graph representation information of all vehicles under dimension \(\mathcal{T}\), constructed from the original states \(\mathcal{S}\) according to graph representation theory, where \((g_1^{\mathcal{T}}, g_2^{\mathcal{T}}, \ldots, g_i^{\mathcal{T}}, \ldots, g_n^{\mathcal{T}}) \in \mathcal{G}_{\mathcal{T}}\); \(\mathcal{A}_{\mathcal{T}}\) represents all possible joint actions \((a_L, a_F) \in \mathcal{A}_{\mathcal{T}}\), where \((A_1^{\mathcal{T}}, A_2^{\mathcal{T}}, \ldots, A_m^{\mathcal{T}}) \in \mathcal{A}_{\mathcal{T}}\);   \(P\left( (g_L', g_H') | (g_L, g_H), (a_L, a_H) \right) \in \mathcal{P}\) is the probability of transitioning from joint graph observation state \(g_{\mathcal{T}}\) to \(g_{\mathcal{T}}'\) when joint action \(a_{\mathcal{T}}\) occurs; and \(\mathcal{R}_{\mathcal{T}}(g_{\mathcal{T}}, a_{\mathcal{T}})\) is the reward calculated when joint action \(a_{\mathcal{T}}\) occurs in joint graph observation state \(g_{\mathcal{T}}\) under decision dimension \(\mathcal{T}\), and can be computed as  \(\mathcal{R}_{\mathcal{T}}(g_{\mathcal{T}}, a_{\mathcal{T}}) = \sum_{i=1}^m r_i^{\mathcal{T}}(g_i^{\mathcal{T}}, a_i^{\mathcal{T}})\), where \(g_i^{\mathcal{T}}\) and \(a_i^{\mathcal{T}}\) are the local graph observation state and action pertaining to the \(i\)-th CAV, respectively. The objective is to learn a joint policy \(\pi_{\mathcal{T}}\) that maximizes the global value function \(Q_{\mathcal{T}}^\pi (g_{\mathcal{T}}, a_{\mathcal{T}}) = E_\pi \left[\sum_{t=0}^{\infty} \gamma^t \mathcal{R}_{\mathcal{T}}(g_t^{\mathcal{T}}, a_t^{\mathcal{T}}) | g_0^{\mathcal{T}} = g_{\mathcal{T}}\right]\).
\end{definition}

The AMG-MDP chain is illustrated in Figure \ref{fig.3}. At  time-step \(t\), the current state of the environment \(s_t\) is observed through graph representation theory, yielding the lane-changing and following dimension graph observation states \(g_L\) and    \(g_H\), respectively. Using these  as inputs, the corresponding  policy networks \(\pi_t^L\) and \(\pi_t^H\) output   lane-changing action \(a_t^L\) and   acceleration action \(a_t^H\), which are input to the state transition probability \(P\) to update the state. Here, the lane-changing action \(a_t^L\) is part of the following dimension graph observation state \(g_H\). Subsequently, rewards \(r_t^L\) and \(r_t^H\) are calculated through each dimension's reward function, and used to update the corresponding dimension's policy networks \(\pi_t^L\) and \(\pi_t^H\). In subsequent Markov decision processes at each time-step, the following dimension is updated at every step, while the lane-changing dimension is updated every \(\Delta\) time-steps.

\subsubsection{State Space}
The state space \(\mathcal{S}_{\mathcal{T}}\) includes the state information of all vehicles in the scenario. The state information for a single vehicle \(s_i^{\mathcal{T}}\) includes data about the self-state, environmental information under dimension \(\mathcal{T}\), and additional information. We  can write
\begin{equation}\label{equ:state}
\left\{ 
\begin{array}{l}
s_i^{\mathcal{T}} = [s^\text{self}, s_{\mathcal{T}}^\text{env} s_{\mathcal{T}}^\text{add}] \\
g_i^{\mathcal{T}} = G_\text{OBS}(s_i^{\mathcal{T}})
\end{array} 
\right.,
\end{equation}
where \(s^{self}\) is the current state information of the vehicle, \(s_{\mathcal{T}}^{env}\) is its    relative information in the environment under dimension \(\mathcal{T}\),   \(s_{\mathcal{T}}^\text{add}\) indicates additional information about the vehicle under dimension \(\mathcal{T}\), and \(G_{OBS}( \cdot )\) is the graph observation function for state information.

Specifically,
\begin{equation}
s^\text{self} = [v, y, l],
\end{equation}
where \(v\) is the vehicle's speed in the direction of travel, \(y\) is its longitudinal position along the lane, and \(l\) is the lane number occupied by the vehicle.

 \(s_{\mathcal{T}}^{env}\) is the current vehicle's safety state with respect to nearby vehicles, helping CAVs to quickly master decisions in the safety domain. For this purpose, we adopt a vehicle risk assessment model based on the safety distance metric between vehicles. Initially, the braking process is divided into four stages to calculate the safety distance, including  communication delay,  braking gap,  linear braking, and  constant braking, and the braking safety distance is
\begin{equation}
D_{\text{sf}} = (t_1 + t_2)v_e + \frac{t_3}{2}(v_e - v_f) + \frac{v_e^2 - v_f^2}{2a_{\max}} + d,
\end{equation}
where \(t_1\) is the sensor delay time during the communication delay, typically ranging from 0.1 s to 0.3 s, with the maximum value of 0.3 s taken in this paper; \(t_2\) is the time to eliminate the braking gap, with a value of 0.1 s; \(t_3\) is the braking time during   linear braking, with a value of 0.24 s;   \(v_e\) is the initial speed of the vehicle; \(v_f\) is its final speed;  \(a_{\max}\) is the maximum deceleration   during braking; and \(d\) is the safety gap between two vehicles after braking to a stop, with a value of 0.6 m.

The actual distance between two vehicles is \(D_{es}\), and the standard safety distance coefficient between vehicles \(i\) and   \(j\) is defined as
\begin{equation}
\kappa_{ij} = \min\left(\frac{D_\text{sf}}{D_\text{es}} - 1, 1\right).
\end{equation}

This formula shows that the standard safety distance coefficient is a dimensionless parameter ranging between --1 and 1. A positive risk value  indicates a potential collision if both vehicles continue under their current states.

Based on the proposed standard safety coefficient, taking a three-lane traffic scenario as an example, the lane-changing dimension decision must consider the standard safety distance coefficients of up to six vehicles ahead and behind in the current and adjacent lanes, while the following dimension only considers   vehicles directly ahead and behind in the current lane. Specifically, the $s_{\mathcal{T}}^\text{env}$ is defined as 
\begin{equation}
\left\{
\begin{array}{l}
s_L^\text{env} = \left[ \kappa_{ij}(j = 1,2, \ldots ,6) \right] \\
s_F^\text{env} = \left[ \kappa_{ij}(j = 1,4) \right]
\end{array}
\ \right.,
\end{equation}
where \(j = 1,2, \ldots, 6\) represent the front, left front, right front, rear, left rear and right rear of the ego vehicle, respectively.
% the vehicles ahead and behind in the same lane, the left lane, and the right lane in order.

Additional information   can be represented as
\begin{equation}\label{equ:addstate}
\left\{
\begin{array}{l}
s_L^\text{add} = [I] \\
s_F^\text{add} = [I, \mathcal{A}_L]
\end{array}
\right.,
\end{equation}
where \(I\) indicates the type of vehicle, where \(I = 1\) represents a CAV, and otherwise, it is an HV; and \(\mathcal{A}_L\) represents the output of the lane-changing dimension decision.

\subsubsection{Action Space}

The action spaces for the lane-changing dimension\(\mathcal{A}_L\), as well as for the following dimension \(\mathcal{A}_F\), encompass lane-changing and acceleration commands, providing complete control for CAVs, where
\begin{equation}
\left\{ 
\begin{array}{l}
\mathcal{A}_L = \{ a_L \in [-1, 0, 1] \}, \\
\mathcal{A}_F = \{ a_F \in [a_\text{dec}, a_\text{acc}] \}
\end{array} 
\right.,
\end{equation}
where \(a_L\) signifies lane-changing commands, with --1, 0, 1 representing a change to the left lane, maintaining the current lane, and a change to the right lane, respectively; and \(a_F\) signifies acceleration commands, where \(a_\text{acc}\) denotes   maximum acceleration, and \(a_\text{dec}\) is maximum deceleration.

\begin{figure}[!t]
\centering
\includegraphics[width=3.45in]{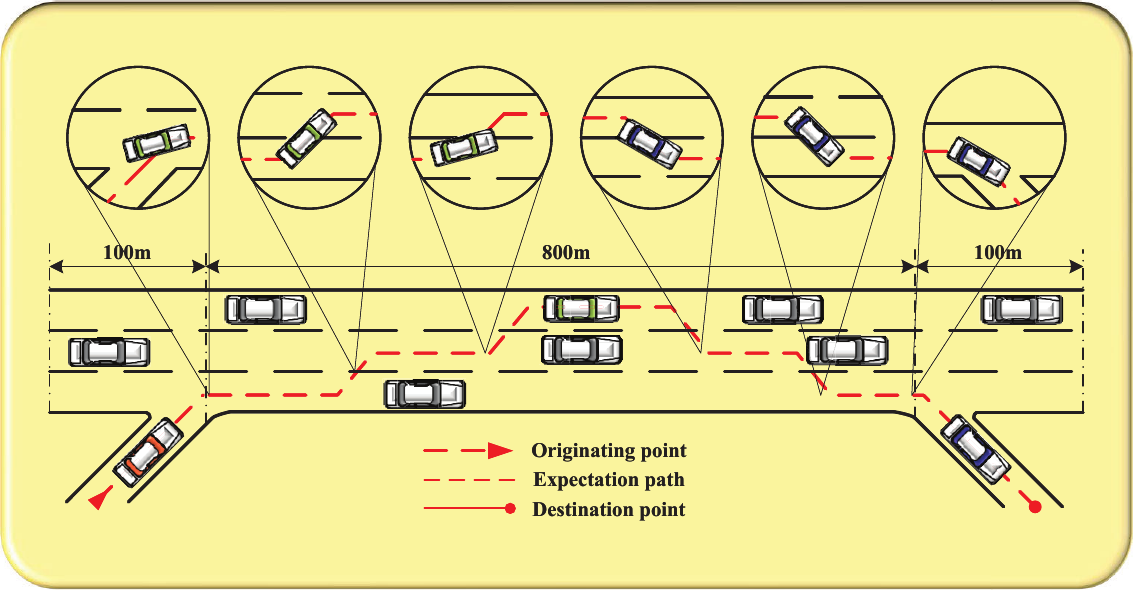}
\caption{Three-lane highway entry/exit scenario.   CAVs enter   main lane from   originating point, change lanes and speed by judging   gap in   traffic flow, and exit from   destination point.}
\label{fig.scene}
\end{figure} 

\subsubsection{Reward Function}
After decomposing the decision system into its respective dimensions, the objectives of the sub-decision modules for each dimension must be examined. The reward functions for both the lane-changing and following dimensions require distinct exploration. We adopt a three-lane highway traffic scenario, where CAVs are set to enter from an on-ramp and exit from the next off-ramp on the same side, as shown in Figure \ref{fig.scene}. For the lane-changing dimension, CAVs must make efficient decisions that align with their destinations. For the following dimension, CAVs are expected to make adaptive decisions that comply with traffic rules and prioritize comfort and energy efficiency.

Initially, the reward function for the lane-changing dimension \(\mathcal{R}_H\) is divided into respective  sub-reward functions for   safety, task, and comfort,
\begin{equation}\label{equ:Lreward}
\left\{ 
\begin{array}{l}
R_L^\text{safe} = - C_1^{\kappa_{ij} + 1}, \quad j = 2,3,5,6, \\
R_L^\text{task} = \left\{ 
\begin{array}{ll}
f_1(L, \Delta y_\text{task}), \\
f_2(L, \Delta y_\text{speed}),
\end{array} 
\right. \\
R_H^\text{comfort} = f_3(\Delta L),
\end{array} 
\right.
\end{equation}
where \(C_1\) is a lane-changing safety constant. \(\kappa_{ij}\) considers only the four vehicles ahead and behind in adjacent lanes. \(L\) denotes the lane occupied by CAVs, corresponding to the fast, middle, and slow lanes. \(\Delta y_\text{task}\) is the longitudinal distance to the exit when in the slow lane. \(\Delta y_\text{speed}\) is the longitudinal distance from the entrance when in the fast lane. The function \(f_1(\cdot)\)  rewards CAVs on the fast lane shortly after entering. The function \(f_2(\cdot)\)  rewards CAVs on the slow lane as they approach the exit. \(\Delta L\) signifies the total number of lane changes by CAVs within a constant time period. \(f_3(\cdot)\) is a linearly increasing function of \(\Delta L\).

The reward function for the following dimension \(\mathcal{R}_L\) is divided into respective  sub-reward functions for safety, task, comfort, and energy,
\begin{equation}\label{equ:Freward}
\left\{ 
\begin{array}{l}
R_F^\text{safe} = - C_2^{\kappa_{ij} + 1}, \quad j = 1,4, \\
R_F^\text{task} = f_4(L, v), \\
R_F^\text{comfort} = f_5(a_L), \\
R_F^\text{energy} = f_6(E),
\end{array} 
\right.
\end{equation}
where \(C_2\) is a following safety constant. \(\kappa_{ij}\) considers only the two vehicles directly ahead and behind in the current lane, and \(v\) is vehicle speed. The function \(f_4(\cdot)\)  rewards CAVs in different lanes if their speed is within a specified range, and penalizes those outside the range, and \(f_5(\cdot)\)  penalizes CAVs whose absolute value of acceleration  exceeds a comfortable range. \(E\) is the energy consumed by CAVs within a time-step. \(f_6(\cdot)\) is a linearly increasing function of \(E\).

\subsection{Parallel Asynchronous Hierarchical Graph Reinforcement Learning}
DRL serves as a pivotal tool to address complex problems. The graphical representations of spatiotemporal interactions, hierarchical learning, and efficient exploration capabilities are methodologies suited for addressing dynamic complexities. Decision-making for CAVs is segmented into lane-changing and following dimensions, which respectively issue high-level commands over intervals of \(\Delta\) time-steps, and  low-level commands at each time-step. Considering the continuity of specific commands, high-level commands utilize an Exponential Greedy Dueling Double Graph Q-Learning Network (EG-GDDQN) to output discrete lane-changing instructions \cite{mnih2015human,wang2016dueling,van2016deep,lillicrap2015continuous}, while low-level commands employ an Exponential Noisy Graph Deterministic Policy Gradient (EN-GDPG) to generate continuous acceleration instructions.

\subsubsection{Exploration Mechanism} 

DRL algorithms enhance environmental exploration by introducing randomness into the actions produced by their policies. Integrating noise into actions   helps to increase exploration efficiency and enhances   the robustness   of the algorithm.   DRL studies have typically employed constant or linear noise for exploration. However, for the commands issued to CAVs, substantial exploration noise is required in the initial stages to thoroughly explore the environment, while reduced exploration noise in later training stages enhances the robustness of decision-making. Linear noise, as demonstrated in prior research, may not adapt well to dynamic environments. Hence, we adopt an Exponential Exploration mechanism for both decision dimensions.

Q-learning is employed for the discrete decision commands in the lane-changing dimension, with exploration based on a greedy algorithm. Within this framework, the multidimensional  DRL actions during   training   can be represented as an
energy sub-reward function \(R_L^{energy}\), where
\begin{equation}\label{equ:Laction}
a_t^L=\left\{\begin{array}{cl}
\underset{a_L \in \mathcal{A}_L}{\arg \max } Q\left(g_t^L, a_L ; \theta_i\right), & p=\left(1-\varepsilon_t\right) \\
U\left(\mathcal{A}_L\right), & p=\varepsilon_t
\end{array},\right.
\end{equation}
where \(U(\cdot)\) signifies the uniform sampling function, and \(\varepsilon_t\) is the exploration rate at time-step \(t\), decreasing over time and calculated based on the initial exploration rate \(\varepsilon_0\),   final exploration rate \(\varepsilon_T\), and   maximum exploration step \(T\) as

\begin{equation}\label{equ:explore}
\varepsilon_t = \varepsilon_0 \exp\left[-\frac{\ln(\frac{\varepsilon_T}{\varepsilon_0})}{T}t\right].
\end{equation}

For   continuous decision commands in the following dimension, the Deterministic Policy Gradient (DPG) method is applied, with exploration achieved by adding noise to   actions,
\begin{equation}\label{equ:Faction}
a_t^F = \text{CLIP}\left(\mu_\omega(g_t^L) + \mathcal{N}_t, a_{\min}^F, a_{\max}^F\right),
\end{equation}
where \(\text{CLIP}(\cdot)\) bounds the final action within the minimum \(a_{\min}^F\) and maximum \(a_{\max}^F\) values. \(\mathcal{N}_t \sim \mathcal{N}(0, \sigma(t))\) represents the noise added to the action at time-step \(t\),  and \(\sigma(t)\) is calculated in a similar manner to Equation \ref{equ:explore}.

Through the adoption of the above behavioral strategies, interactions between CAVs and the environment generate two-dimensional tuples,
\begin{equation}
\left\{ 
\begin{array}{l}
\zeta_t^L = (s_t, g_t^L, a_t^L, r_t^L, s_{t+1}, g_{t+1}^L) \\
\zeta_t^F = (s_t, g_t^F, a_t^F, r_t^F, s_{t+1}, g_{t+1}^F)
\end{array} 
\right.,
\end{equation}
which
 are then stored in the corresponding experience replay buffers containing training data.

\subsubsection{Learning objective}
For   discrete decisions for the lane-changing dimension,   Q-learning   may overestimate the true value, and this overestimation is non-uniform. There are two main reasons for this:  bootstrapping leads to the propagation of biases, and maximization   causes the Temporal Difference (TD) target to overestimate the true value. Therefore, we adopt the Dueling network and Double algorithm. The Dueling Network architecture,   \(\theta_i\), includes an optimal state-value function \(\theta^V\) and   optimal advantage function \(\theta^D\).  Hence, the gradient computation formula for EG-GDDQN based on AMG-MDP is
\begin{equation}\label{equ:Lvalue}
\left\{ 
\begin{array}{l}
Q(g_t^L, a_L; \theta_i) = V(g_t^L; \theta^V) + D(g_t^L, a_L; \theta^D) \\
\quad\quad\quad\quad\quad\quad - \mathop{\rm mean}\limits_{a_L \in \mathcal{A}_L} D(g_t^L, a_L; \theta^D), \\
\delta_i = Q(g_t^L, a_L; \theta_i) - r_t^L - \gamma Q(g_{t+1}^L, a_{t+1}^L; \theta_i^-),
\end{array} 
\right.
\end{equation}
 where  $\mathop{\rm mean}$  addresses the issue of non-uniqueness; in practice, this has been found to perform slightly better than the max operation.
 
For the following dimension's continuous decisions, random policies, due to their action variability, can degrade CAV performance because the update direction of random policy parameters may not align with the optimal direction of policy gradients. Unlike random policies that integrate action and state spaces, deterministic policies only integrate the state space, meaning that for a given state and parameters, only one specific action is output. The Actor gradient \(\nabla_{\omega^\mu} J\) and   loss function for value \(L(\varphi, \mathcal{D}_F)\)   for EN-GDPG based on AMG-MDP are:
\begin{equation}\label{equ:Factor}
\begin{aligned}
\nabla_{\omega^\mu} J & \approx \mathbb{E}_{g_t^L \sim \rho^\beta}\left[\nabla_{\omega^\mu} Q\left(g_L, a \mid \omega^Q\right) \mid g_L=g_t^L,\right. \\
& \quad\left.a=\mu\left(g_t^L \mid \omega^\mu\right)\right] \\
& =\mathbb{E}_{g_t^L \sim \rho^\beta}\left[\nabla_a Q\left(g_L, a \mid \omega^Q\right) \mid g_L=g_t^L,\right. \\
& \quad\left.a=\mu\left(g_t^L\right) \cdot \nabla_{\omega^\mu} \mu\left(g_L \mid \omega^\mu\right)\right]
\end{aligned}
\end{equation}

\begin{equation}\label{equ:Fvalue}
\begin{aligned}
& L\left(\varphi, \mathcal{D}_F\right)=\underset{\left(s_t, \mathrm{~g}_t^F, a_t^F, r_t^F, s_{t+1}, \mathrm{~g}_{t+1}^F\right) \sim \mathcal{D}_F}{\mathbb{E}}\left[\left(Q_{\varphi}\left(g_L, a_F\right)-\right.\right. \\
& \quad\quad\quad\left.\left.\left(r_{t+1}^F+\gamma(1-d) \max _{a_{t+1}^L} Q_{\varphi}\left(g_{t+1}^F, a_{t+1}^F\right)\right)\right)^2\right].
\end{aligned}
\end{equation}
\subsubsection{Integration} 

PAH-GRL, through the integration of the above mechanisms, leverages multilevel graph representation Theory, and is described as Algorithm 1. Given the extensive body of research, this paper omits discussion of   DRL algorithms.
\IncMargin{1em}
\begin{algorithm} \SetKwData{Left}{left}\SetKwData{This}{this}\SetKwData{Up}{up} \SetKwFunction{Union}{Union}\SetKwFunction{FindCompress}{FindCompress} \SetKwInOut{Input}{input}\SetKwInOut{Output}{output}
	
        \Input{ initial $F$-policy parameters $\omega $, $F$-Q-function parameters $\varphi $,$F$-empty replay buffer ${{\cal D}_F}$, $F$-training batch size ${U_{F}}$,$L$-Q-function parameters $\theta $,$L$-empty replay buffer ${{\cal D}_L}$, $L$-training batch size ${U_{L}}$} 
        % \Output{ } 
	\emph{Set target parameters equal to main parameters ${\omega_{targ}} \leftarrow \omega$,${\varphi_{targ}} \leftarrow \varphi$,${\theta_{targ}} \leftarrow \theta$}\;
	 \BlankLine 
	 \For{episode$ \in \{ 0,1, \ldots ,M\} $}{ 
	 	\emph{Based on Equation~(\ref{equ:state}), receive initial observation state $s_0$, graph observation states \(g_0^{L}\) and \(g_0^{F}\)}
            
	 	\For{t $ \in \{ 0,1, \ldots ,T\} $}{\label{forins}

                  \For{t $ \in \{ 0,5, \ldots ,\left\lfloor {T/\Delta} \right\rfloor $\} }{
                  \emph{Based on Equation~(\ref{equ:Faction}), select action \(a_t^{F}\) }
                  \emph{Obtain reward \(r_t^{F}\), new state $s_{t+1}$ and \(g_{t+1}^{F}\) }

                  \emph{Add tuple ${ ({s_t},{\rm{g}}_t^F,a_t^F,r_t^F,{s_{t + 1}},{\rm{g}}_{t + 1}^F)}$ into ${{\cal D}_F}$ }

                  \emph{Sample ${U_{F}}$ tuples from ${{\cal D}_F}$ }

                  \emph{Update $F$-value network in Eqn.~(\ref{equ:Fvalue}})

                  \emph{Update $F$-actor network in Eqn.~(\ref{equ:Factor}})

                  \emph{Update $F$-target networks ${\omega_{targ}} \leftarrow \omega$,${\varphi_{targ}} \leftarrow \varphi$}

                  \emph{Update graph observation states \(g_0^{L}\) in Eqn.~(\ref{equ:addstate}})

                  }
            \emph{\textbf{end}}
            
            \emph{Based on Eqn.~(\ref{equ:Laction}), select action \(a_t^{L}\)}
            
            \emph{Obtain reward \(r_t^{L}\), new state $s_{t+1}$ and \(g_{t+1}^{L}\) }

            \emph{Add tuple ${ ({s_t},{\rm{g}}_t^L,a_t^L,r_t^L,{s_{t + 1}},{\rm{g}}_{t + 1}^L)}$ into ${{\cal D}_L}$ }

            \emph{Sample ${U_{L}}$ tuples from ${{\cal D}_L}$ }

            \emph{Update $L$-value network in Eqn.~(\ref{equ:Lvalue}})

            \emph{Update $L$-target value network ${\theta_{targ}} \leftarrow \theta$}
 	 	   }
        \emph{\textbf{end}}
 		 }
     \emph{\textbf{end}}
   
 	 	  \caption{PA-HGRL}
 	 	  \label{algo_disjdecomp} 
 	 \end{algorithm}
 \DecMargin{1em} 

% \begin{figure}[!t]
% \centering
% \includegraphics[width=3.5in]{network}
% \caption{ Three-lane high-speed road entry/exit scenario. the CAVs enter the main lane from the originating point, change lanes and speed by judging the gap in the traffic flow, and exit from the destination point.}
% \label{fig.scene}
% \end{figure} 

\subsection{Multilevel Multi-head Graph Attention Network Module}  \label{section4.3}
The construction of the graph network model must be compatible with the matrix of the previously proposed multilevel graph representation. Thus, a Multilevel Multi-head Graph Attention Network (ML-MGAT) is introduced to better handle feature information across different dimensions.

Specifically, for a target node \(v\), \(\mathcal{N}_u\) represents all neighboring nodes of \(u\). For each pair of nodes \( (u, v) \), the normalized attention coefficient \( e_{ij}^k \) is computed only when \( a_{uv} \) is nonzero (i.e., there exists an edge between nodes \(u\) and \(v\)), ensuring that attention coefficients are calculated only between actually connected nodes \cite{veličković2018graph}. For dimension \( \mathcal{T} \), the revised formula for   the normalized attention coefficients is

\begin{equation}
\alpha _{u,v}^{\cal T} = \frac{{\exp \left( {{\mathop{\rm LeakyReLU}\nolimits} \left( {{a^{T,{\cal T}}} \cdot \left[ {{W^{\cal T}}h_u^{\cal T}{W^{\cal T}}h_v^{\cal T}} \right]} \right) \cdot a_{uv}^{\cal T}} \right)}}{{\sum\limits_{u \in {\cal N}_u^{\cal T}} {\exp } \left( {{\mathop{\rm LeakyReLU}\nolimits} \left( {{a^{T,{\cal T}}} \cdot \left[ {{W^{\cal T}}h_u^{\cal T}{W^{\cal T}}h_v^{\cal T}} \right]} \right) \cdot a_{uv}^{\cal T}} \right)}},
\end{equation}
where \(W^{\cal T}\) is a learnable projection matrix,   \(a^{T,{\cal T}}\) is a trainable vector indicating the direction of attention, and \( [ \cdot || \cdot ] \) denotes  vector concatenation.

A multi-head attention mechanism is introduced, where  heads can learn different  information,

\begin{equation}
h_u^{\cal T} = \sum\limits_{k = 1}^K \sigma  \left( {\sum\limits_{v \in {\cal N}_u^{\cal T}} {\alpha _{u,v}^{\cal T}} {W^{\cal T}}h_v^{\cal T}} \right),
\end{equation}
where \(\sigma\) is a nonlinear activation function, and \(K\) is the number of heads.

  ML-MGAT allows the model to more finely consider the relationships between nodes, such as by adjusting the allocation of attention based on factors like relative distance and speed between vehicles, so the multi-head graph attention network can more precisely capture the dynamic changes and multidimensional interactions in the traffic network, providing deeper and more detailed information support for autonomous driving decisions. The policy network in the Dueling network should be replaced with   state value  and   action advantage sub-networks. The value network in EN-GDPG has a structure similar to the policy network, except that the action variable is added as an input, and the output is replaced by Q-values. We set the value network to be wider than the policy network.

\section{ Experiments}

We evaluated the performance of the proposed spatiotemporal interaction-enhanced decision-making framework. We introduce the experimental platform, explain the parameters of the designed   heterogeneous mixed autonomy traffic scenario, and describe the baseline methods and testing metrics.
\begin{table}[!t]
\caption{Experimental platform parameter settings.\label{tab:table1}}
\newcolumntype{C}{>{\centering\arraybackslash}X}
\centering
\begin{tabular}{cc}
\toprule

\textbf{Parameter} & \textbf{Value}      \\
\midrule

Operating system & Ubuntu 20.04\\

Language  & Python 3.7.3\\

CUDA version & CUDA 11.3\\

PyTorch  & v1.14.0\\

torch-geometric  & v2.0.4\\

Optimizer & Adam\\

Nonlinearity  & ReLU\\
Number of vehicles, $n$& 50\\
Number of CAVs, $m$& 5\\
\multirow{3}{*}{Traffic density of HVs} &  lane0:  360 ${\rm{vehicles/h}}$\\
 & lane1: 720 ${\rm{vehicles/h}}$\\
  & lane2: 720 ${\rm{vehicles/h}}$\\
\multirow{3}{*}{Initial velocity of HVs} &  lane0: 12 ${\rm{m/s}}$\\
 & lane1: 18 ${\rm{m/s}}$\\
  & lane2: 22 ${\rm{m/s}}$\\
Initial velocity of CAVs & 15 ${\rm{m/s}}$\\
Longitudinal speed limitation  & 35 ${\rm{m/s}}$\\
Length of vehicle  & 5 ${\rm{m}}$\\

\hline
\end{tabular}
\end{table}
\subsection{Experimental Platform} 

Simulations were conducted on a computer equipped with a standard keyboard. The hardware supporting real-time computations included an Intel Core i7-11800H CPU @ 2.30 GHz, with 16 GB RAM, and an NVIDIA GeForce RTX 3090 GPU.  Flow \cite{Wu_2022} was utilized, with algorithms and scripts programmed in Python. Neural network models were constructed using the PyTorch framework. The hyperparameters related to this study are shown in Table \ref{tab:table1}.

\subsection{ Simulation Scenario Setup}

High-density, high-randomness, and high-dynamic highway traffic scenarios are typical of heterogeneous mixed autonomy. The primary challenges CAVs face in operating efficiently and safely include: 1) minimizing   disruption to overall traffic flow when entering and exiting the highway; 2) making complex multi-objective decisions while sharing the road with HVs, considering the surrounding vehicles' speed and position, traffic flow density, potential conflicts, passenger comfort, and energy consumption requirements; 3) dynamic and continuous spatiotemporal propagation of decisions made by CAVs; e.g., current driving behaviors can be transmitted from HVs to subsequent CAVs.

We designed a 1-kilometer, three-lane highway section, incorporating practical tasks and requirements, as shown in Figure \ref{fig.scene}. Colored vehicles represent autonomous cars, which   entered the scenario from the left junction and exited from the right junction. In this interactive scenario, CAVs had to execute four logical tasks: choosing the right time to merge onto the main road, lane-changing to the fast lane and accelerating, lane-changing to the slow lane and decelerating, and choosing the appropriate opportunity to exit. We established the following rules  according to   actual traffic scenarios and computational costs:
\begin{enumerate}[(a)]
\item	Each episode was limited to 150 seconds, with five CAVs randomly appearing within 50 seconds, ensuring their interaction;
\item	HVs (grey vehicles)  appeared randomly on roads, overlapping CAVs' routes with a certain traffic density;
\item   CAVs could continue to drive on the right main road, but this was considered a mission failure;
 \item	CAVs could always drive on the slow lane and exit, but this strategy was considered as low-scoring;
\item	Lanes had different speed ranges, and speeding or driving too slowly were considered to be violations;
\item   All vehicles had to strictly adhere to traffic rules.
\end{enumerate}

The  above rules were established to closely adapt to real traffic environments. HVs on different lanes pose traffic flow challenges to CAVs, which must choose areas with low traffic flow density for lane-changing and speed adjustment, minimizing the impact on HVs. Scenario parameters are given in Table \ref{tab:table1}.

\begin{table}[!t]
\caption{Comparison of    algorithms.\label{tab:table2}}
\newcolumntype{C}{>{\centering\arraybackslash}X}
\centering
\begin{tabular}{ccccc}
\toprule
\textbf{Algorithm} & \textbf{L-dimension}    & \textbf{H-dimension}  & \textbf{Network} &  \textbf{Graph}  \\
\midrule
D3QN &\multicolumn{2}{c}{D3QN} &DNN & None\\
PAH-DNN &EG-D3QN&EN-DDPG &DNN & None\\
PAH-GCN &EG-GDDQN&EN-GDPG &GCN & Basic\\
PAH-GAT &EG-GDDQN&EN-GDPG &MGAT & Basic\\
PAH-MLGCN &EG-GDDQN&EN-GDPG &GCN & Multi-level\\
PAH-MLGAT &EG-GDDQN&EN-GDPG &MGAT & Multi-level\\
\hline
\end{tabular}
\end{table}
\begin{table}[!t]
\caption{Hyperparameter  settings.\label{tab:table3}}
\newcolumntype{C}{>{\centering\arraybackslash}X}
\centering
\begin{tabular}{ccc}
\toprule

\textbf{Parameter} & \textbf{L-dimension}    & \textbf{F-dimension}  \\
\midrule
\multirow{2}{*}{Learning rate}   & \multirow{2}{*}{0.00075}   & Critic: 0.0001\\
 & & Actor: 0.000025\\
Time-step & 0.5 s & 0.1 s\\
Starting exploration value  & 0.6 & 0.5\\
Terminating exploration value  & 0.02 & 0.05\\
Continuous exploration time-step  & 140000 & 700000\\
Mini batch size & 64& 64\\
Discount rate & \multicolumn{2}{c}{0.99}\\
Time-step of   current network update & 100& 200\\
Time-step of   target network update & 400& 800\\
Soft update rate & 0.01 & 0.005\\
Number of heads, $K$& 3 & 3\\
% Updating rate  & 0.05\\

% Minibatch size & 64\\

%  Starting greedy rate & 0.4\\
%  Ending greedy rate & 0.0075\\
%  Time step of greedy algorithm duration & 100000\\

% Time step  & 0.1 ${\rm{s}}$\\

% Traffic density of HVs. & 1700 ${\rm{vehicles/h}}$\\

% The initial velocity of HVs & 15 ${\rm{m/s}}$\\

% The initial velocity of CAVs & 15 ${\rm{m/s}}$\\
\hline
\end{tabular}
\end{table}
\subsection{Baseline Methods}

We employed EN-GDPG, EG-GDDQN, GCN, MGAT, and other methods, embedded in the proposed framework for comparison through ablation and contrast experiments. Six   algorithms are shown in Table \ref{tab:table2}, whose hyperparameters   are given  in Table \ref{tab:table3}. We emphasize several points:
\begin{enumerate}[(a)]
\item	The basic graph representation theory (section \ref{section:3.1}) is considered a baseline comparison, and multilevel graph representation theory (section \ref{section:3.2}) is used for further comparison;
\item	An ablation experiment training results graph is provided for the last four algorithms in Table \ref{tab:table2}. All algorithms were trained with five different random seeds, and results   compared using mean and variance;
\item   All algorithms were compared in  tests. Each model was tested 10 times, and all the test indicators are average values. 
% with the mean taken in the end.
\end{enumerate}

Reward values, exit rates, emergency braking incidents, task time, and energy consumption were used  to assess the    performance, driving efficiency, driving comfort, and energy efficiency of the all algorithms in Table \ref{tab:table2}. A trajectory analysis was conducted based on the medians of the test rewards.
\begin{figure}[t!]
    \centering
    
    \subfigure[Lane-change dimension.]{
        \begin{minipage}[t]{0.9\linewidth}
            \includegraphics[width=8cm]{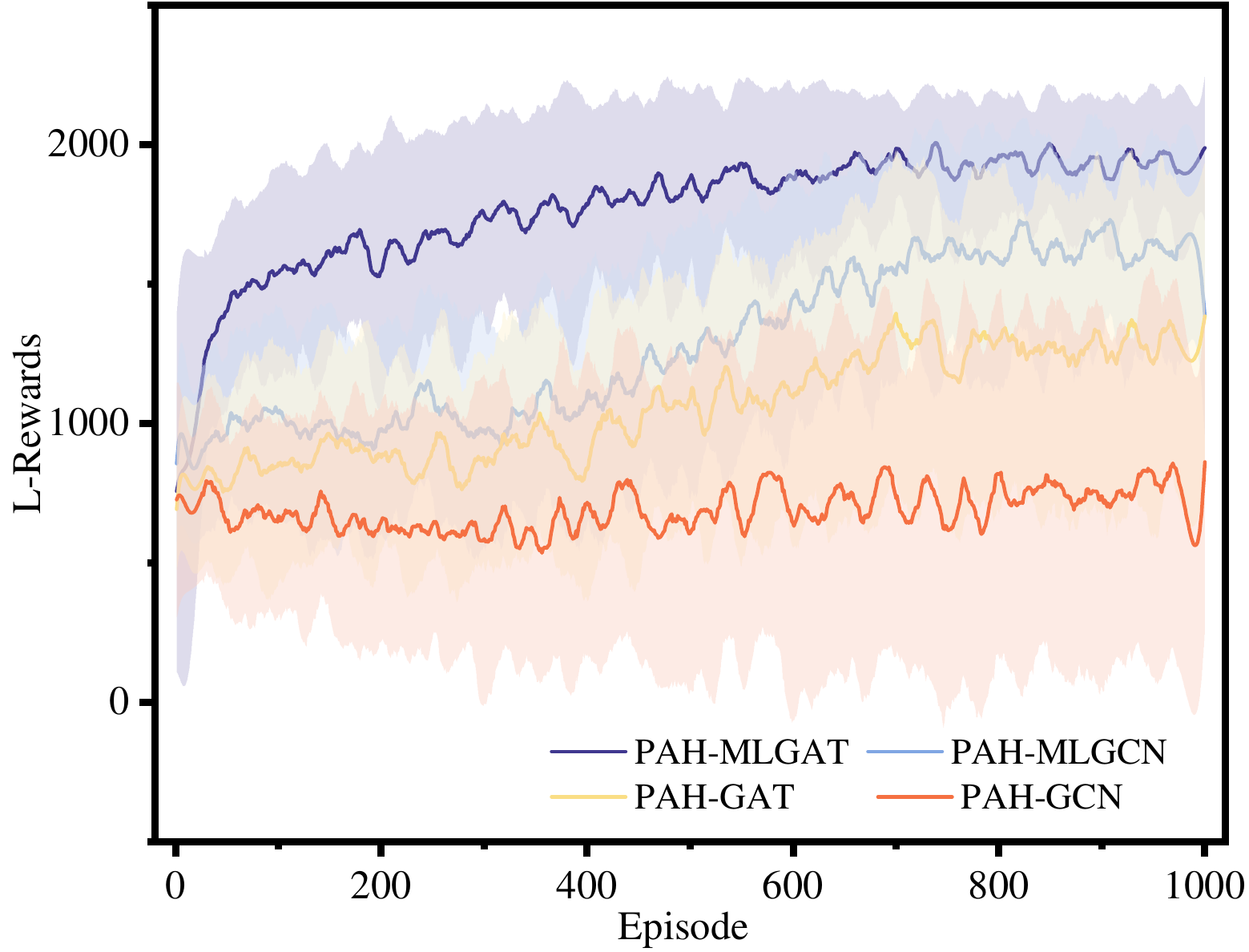}
            
        \end{minipage}
    } 
    \subfigure[Following dimension.]{
        \begin{minipage}[t]{0.9\linewidth}
            \includegraphics[width=8cm]{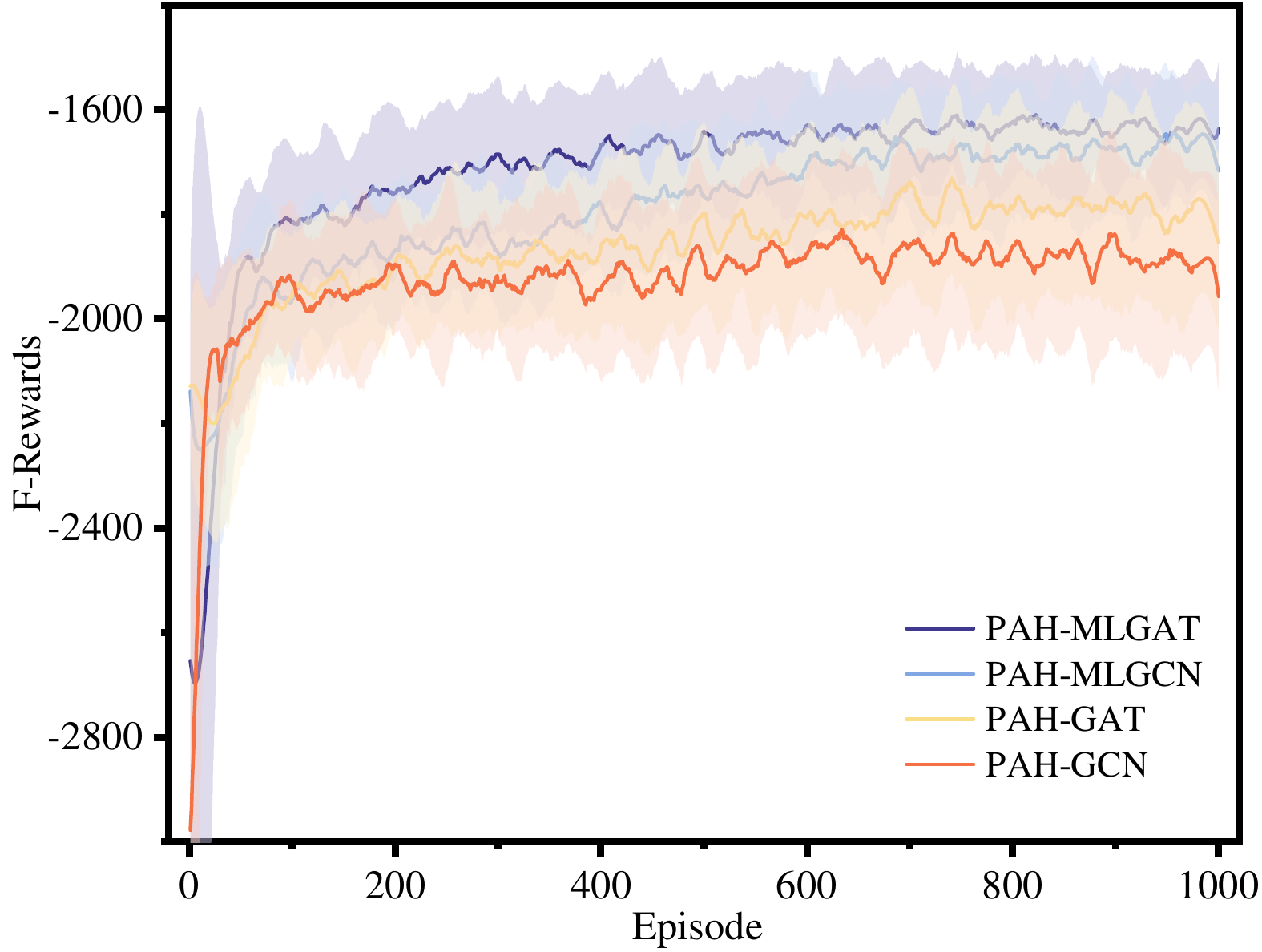}
            
        \end{minipage}
    }
    
    \caption{Variations in   rewards of   two decision-making dimensions over       1000 episodes in ablation training experiments.}
    \label{fig:reward}
\end{figure}

% \begin{figure*}[bt!]
%     \centering
    
%     \subfigure[The corresponding speed in pedestrian protection traffic scenario.]{
%         \begin{minipage}[t]{0.315\linewidth}
%             \includegraphics[width=5.5cm]{NUMBER.eps}
            
%         \end{minipage}
%     } 
%     \subfigure[The corresponding speed in the single-attribute pedestrian traffic scenario.]{
%         \begin{minipage}[t]{0.315\linewidth}
%             \includegraphics[width=5.5cm]{TIME.eps}
            
%         \end{minipage}
%     }
%     \subfigure[The corresponding speed in the complex-attributes pedestrian traffic scenario.]{
%         \begin{minipage}[t]{0.315\linewidth}
%             \includegraphics[width=5.5cm]{ENERGY.eps}
            
%         \end{minipage}
%     }
%     \caption{Corresponding trajectories and vehicle speeds for the episode with optimal reward value during multi-modal simulation testing.}
%     \label{fig:index}
% \end{figure*}
\begin{figure}[!t]
\centering
\includegraphics[width=3.6in]{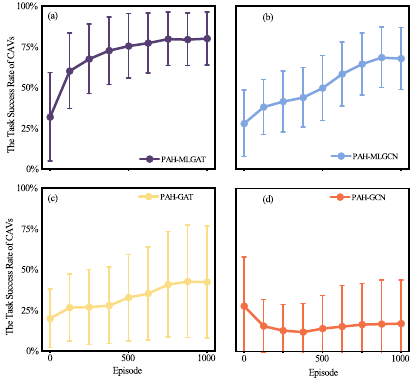}
\caption{Task success rate of CAVs of different GRL methods: (a) PAH-MLGAT; (b) PAH-MLGCN; (c) PAH-GAT; (d) PAH-GCN. Error bar refers to   standard deviation.}
\label{fig.task}
\end{figure} 

\begin{figure}[!t]
\centering
\includegraphics[width=3.6in]{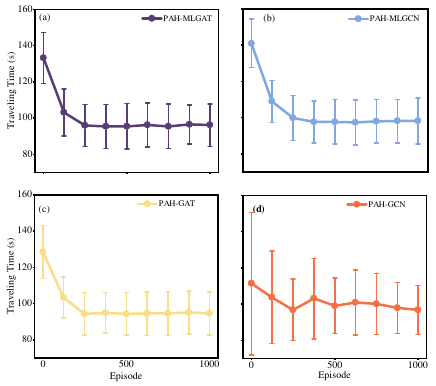}
\caption{Traveling time of CAVs of different GRL methods: (a) PAH-MLGAT; (b) PAH-MLGCN; (c) PAH-GAT; (d) PAH-GCN. Error bar refers to   standard deviation.}
\label{fig.time}
\end{figure} 

\begin{figure}[!t]
\centering
\includegraphics[width=3.6in]{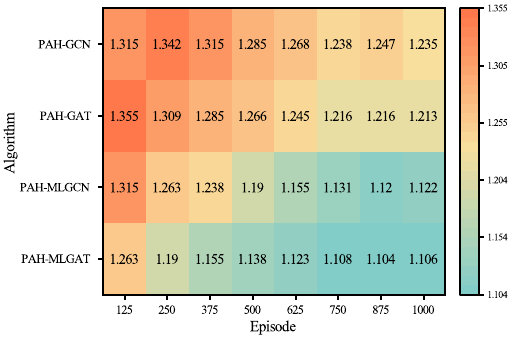}
\caption{Energy consumption of   ablation study. Quantities are energy expenditures of CAVs during   training phase of   proposed algorithm.}
\label{fig.energy}
\end{figure} 

\section{Results}
We analyze the training process of the ablation experiments. Also, we conducted a comparative test experiment involving six algorithms, followed by a comprehensive comparison. Finally, we present an analysis of the spatiotemporal trajectories of the optimal model.
\subsection{Ablation Training Experiments}
Figure \ref{fig:reward} displays the reward values during each training iteration of the four algorithms in   ablation experiments. Within the proposed PAH-GRL framework, both the proposed multilevel graph representation  and the corresponding MGAT exhibited superior performance compared to the traditional GCN algorithm. PAH-GCN showed continually expanding error during   training, indicating that the basic graph representation struggles to precisely represent the interactive behavior of heterogeneous mixed autonomy. The introduction of the multilevel graph representation module resulted in a smoother increase in the reward curve during training, suggesting  effective extraction of scenario features, enhancing the learning capabilities of CAVs.   MGAT based on multilevel graph representation also displayed higher rewards, indicating improved exploration abilities for CAVs. PAH-MLGAT demonstrated the optimal convergence speed and highest exploration rewards in both decision dimensions. Importantly, even with lane-changing dimension outputs serving as inputs for the following dimension, rewards in both dimensions continued to rise steadily. This validates the   robustness and stability of the proposed PAH-GRL framework.

Figures \ref{fig.task}-\ref{fig.energy} respectively illustrate the task success rate, total driving time, and energy consumption for each training iteration of the four algorithms in   ablation experiments. As shown in Figure \ref{fig.task}, the task success rates of the four algorithms align with the trend of reward values in the lane-changing dimension, indicating that PAH-MLGAT better assesses the motivation for lane changes. However, the final success rate did not meet expectations due to the residual randomness in the greedy algorithm used in the lane-changing dimension toward the end of training. A more precise analysis of the task success rate can be obtained in the test experiments if greedy actions are omitted. As shown in Figure \ref{fig.time}, there was no significant difference in the total time per episode among the four algorithms, as only the exit of the last CAV concluded the episode, making it difficult to reflect overall driving efficiency. Therefore, we analyzed the variance in driving time, and supplemented it with a CAV trajectory analysis. In the figure, PAH-GCN exhibited large fluctuations in driving time, while PAH-GAT had shorter   times. However, trajectory analysis revealed frequent speed changes by CAVs in the slow lane in both cases, reducing mission time. From the energy analysis in Figure \ref{fig.energy}, PAH-GCN and PAH-GAT experienced frequent acceleration changes, increasing  energy consumption. A comprehensive analysis of these three   training indicators shows that PAH-MLGAT-trained CAVs can effectively aggregate other vehicles' driving characteristics and interaction topologies, enhancing traffic efficiency and decision robustness while reducing  energy consumption.
\begin{table*}
\caption{Testing evaluation results.\label{tab:table4}}
\newcolumntype{C}{>{\centering\arraybackslash}X}
\centering
\begin{tabular}{ccccccccccccc} 
\hline
\centering
\multirow{2}{*}{\textbf{Algorithms}} & \multicolumn{2}{c}{${p_{success}}$} &  \multicolumn{2}{c}{${N_{braking}}$} &\multicolumn{2}{c}{${T_{travel}}$} & \multicolumn{2}{c}{$Energy$}  & \multicolumn{2}{c}{ L-Reward} & \multicolumn{2}{c}{ F-Reward}\\
                 &$\mu (s)$ &$\sigma $                   & $\mu $ &$\sigma $                  &$\mu $  &  $\sigma $                 & $\mu(J) $ &  $\sigma $                & $\mu $ &$\sigma $  & $\mu $ &$\sigma $\\ 
\hline
\specialrule{0em}{2pt}{1pt}
\centering
    D3QN              & 10.40\%							 &      0.098
    & 7.560 & 1.215                  &  1282.51& 152.03                  &1.577  &   0.211   & -  &   -            & -  & -    \\
\specialrule{0em}{1pt}{0pt}
     PAH-DNN								             & 13.60\% &   0.135               &5.920  &                1.383   & 1179.60 &  134.60                & 1.301 &0.116      & 0.3842 &0.1591  & 0.3155 &0.1356             \\    
\specialrule{0em}{1pt}{0pt}
     PAH-GCN								             & 20.80\% &   0.132               &5.160  &                0.967   & 1004.28 &  131.84                & 1.198 &0.1629      & 0.4689 &0.1338    & 0.3558 &0.1406             \\
\specialrule{0em}{1pt}{0pt}
     Imp. $\uparrow$								             & 52.94\% &   \underline{2.22\%}                &12.84\%  &                30.08\%   & 14.96\% &  2.05\%                 & 7.92\% &-40.43\%       & 22.04\% &15.90\%           & 12.77\% &-3.69\%       \\
\specialrule{0em}{1pt}{0pt}
                 PAH-GAT						 &54.40\%  &          0.223         & 3.360 &                	0.889   &951.96  &               	129.46    & 1.219 &  0.152 & 0.7294 &  0.1043   & 0.5011 &  0.1272                 \\
\specialrule{0em}{1pt}{0pt}
     Imp. $\uparrow$								             & 300.00\% &   -65.18\%                &43.24\%  &                35.72\%   & \textbf{19.30\%} &  3.82\%                 & 6.30\% &-31.03\%       & 89.85\% &34.44\%           & 58.82\% &\underline{6.20\%}       \\
\specialrule{0em}{1pt}{0pt}
                  PAH-MLGCN						 &74.40\%  &          0.174         & 2.320 &                	0.733   &971.72  &               	114.99    & 1.137 &  0.1241 & 0.8178 &  0.0656   & 0.6630 &  0.1323               \\
\specialrule{0em}{1pt}{0pt}
     Imp. $\uparrow$								             & \underline{447.06\%} &   -28.89\%                &\underline{60.81\%}  &                \textbf{47.00\%}   & 17.62\% &  \underline{14.57\%}                 & \underline{12.61\%} &\underline{-6.98\%}       & \underline{112.86\%} &\underline{58.77\%}           & \underline{110.14\%} &2.43\%       \\
\specialrule{0em}{1pt}{0pt}
                  PAH-MLGAT					 &95.20\%  &          0.085         & 1.320 &                	0.835   &960.52  &               	103.24    & 1.0751 &  0.0821 & 0.9063 &  0.0376        & 0.7963 &  0.0910            \\
\specialrule{0em}{1pt}{0pt}
     Imp. $\uparrow$								             & \textbf{600.00\%} &   \textbf{37.03\% }               &\textbf{77.70\%}  &                \underline{39.62\%}  & \underline{18.57\%} &  \textbf{23.30\% }                & \textbf{17.36\%} &\textbf{29.22\%}       & \textbf{135.89\%} &\textbf{76.37\%}           & \textbf{152.39\%} &\textbf{32.89\%}       \\                  
% \specialrule{0em}{1pt}{0pt}
%      Imp. $\uparrow$								             & 14.71\% &   30.57\%                &\underline{56.85\%}  &                48.65\%   & 44.74\% &  40.46\%                 & 4.45\% & 29.85\%      & 72.86\% &15.04\%

\specialrule{0em}{1pt}{1pt}
\hline
\end{tabular}
\begin{tablenotes}
\item Improvements for each algorithm compared with   PAH-DNN   are shown as Imp. Best results are in boldface; second-best are underlined.

\end{tablenotes}
\end{table*}
\subsection{Comparative Test Experiment}
Table \ref{tab:table4} shows that PAH-MLGAT achieved the best overall performance and robustness across six testing metrics in heterogeneous mixed autonomy. This demonstrates that PAH-MLGAT effectively couples multilevel graph representation, PAH-GRL, and multi-head graph attention mechanisms, significantly enhancing the comprehensive performance and robustness of CAVs' decisions. In contrast, D3QN, which outputs both lane-changing and acceleration commands, struggled to facilitate normal driving in heterogeneous mixed autonomy, and exhibited high variance in performance metrics. The introduction of Parallel Asynchronous Hierarchical and basic graph representation theory gradually enabled CAVs to achieve set driving goals. However, although PAH-GCN and PAH-GAT reduced the time consumed per round, they failed to master the decision strategy for exiting the highway and conserving energy. Furthermore,   multilevel graph representation aided CAVs in effectively capturing the spatiotemporal dependencies of scenarios, leading to higher driving efficiency, comfort, energy conservation, and robustness.

% \clearpage
\begin{figure*}[bt!]
    \centering
    \subfigure[3D spatiotemporal trajectory test diagram of PAH-MLGAT.]{
        \begin{minipage}[t]{0.315\linewidth}
            \includegraphics[width=5.5cm]{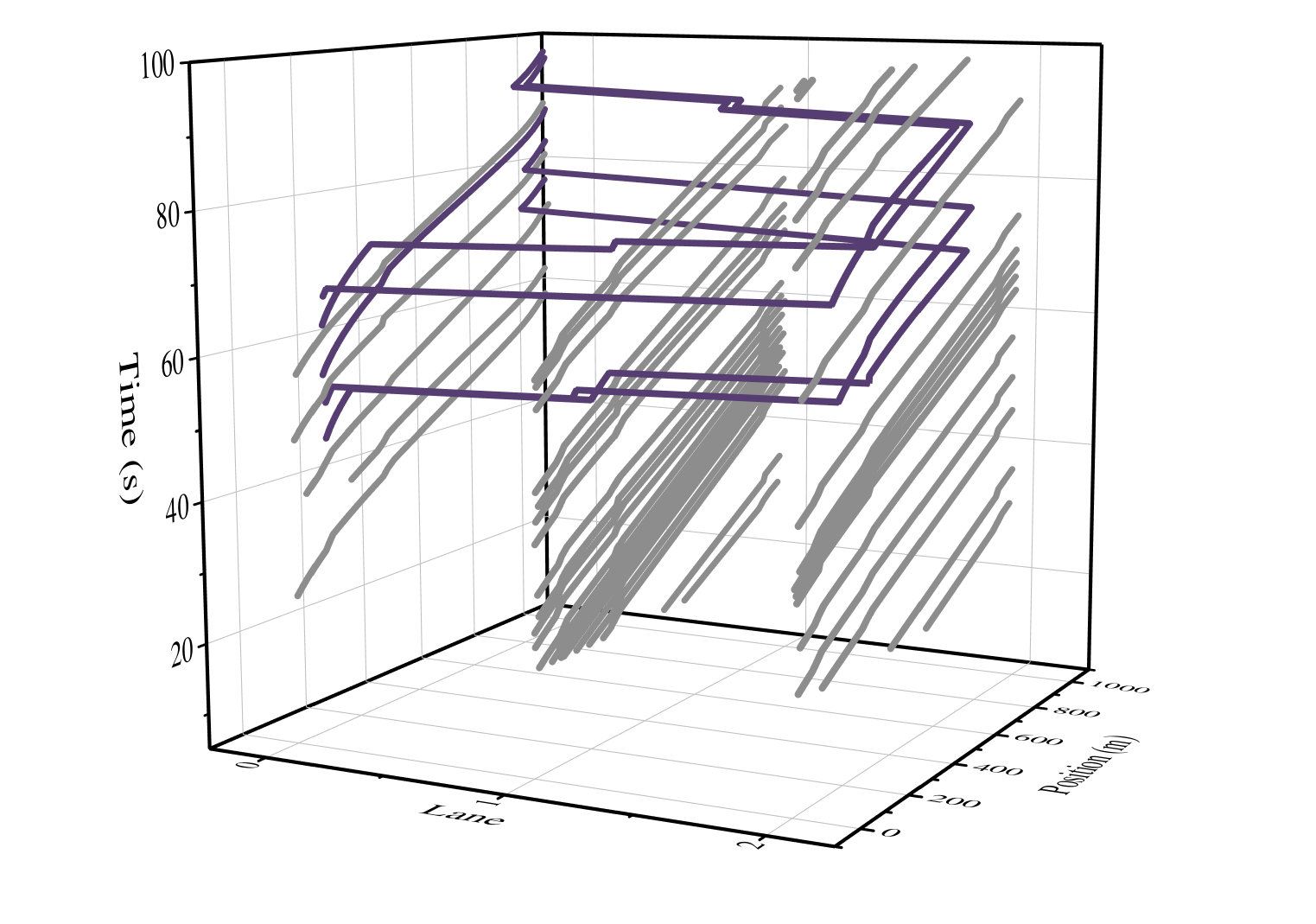}
            
        \end{minipage}
    } 
    \subfigure[3D spatiotemporal trajectory test diagram of PAH-GAT.]{
        \begin{minipage}[t]{0.315\linewidth}
            \includegraphics[width=5.5cm]{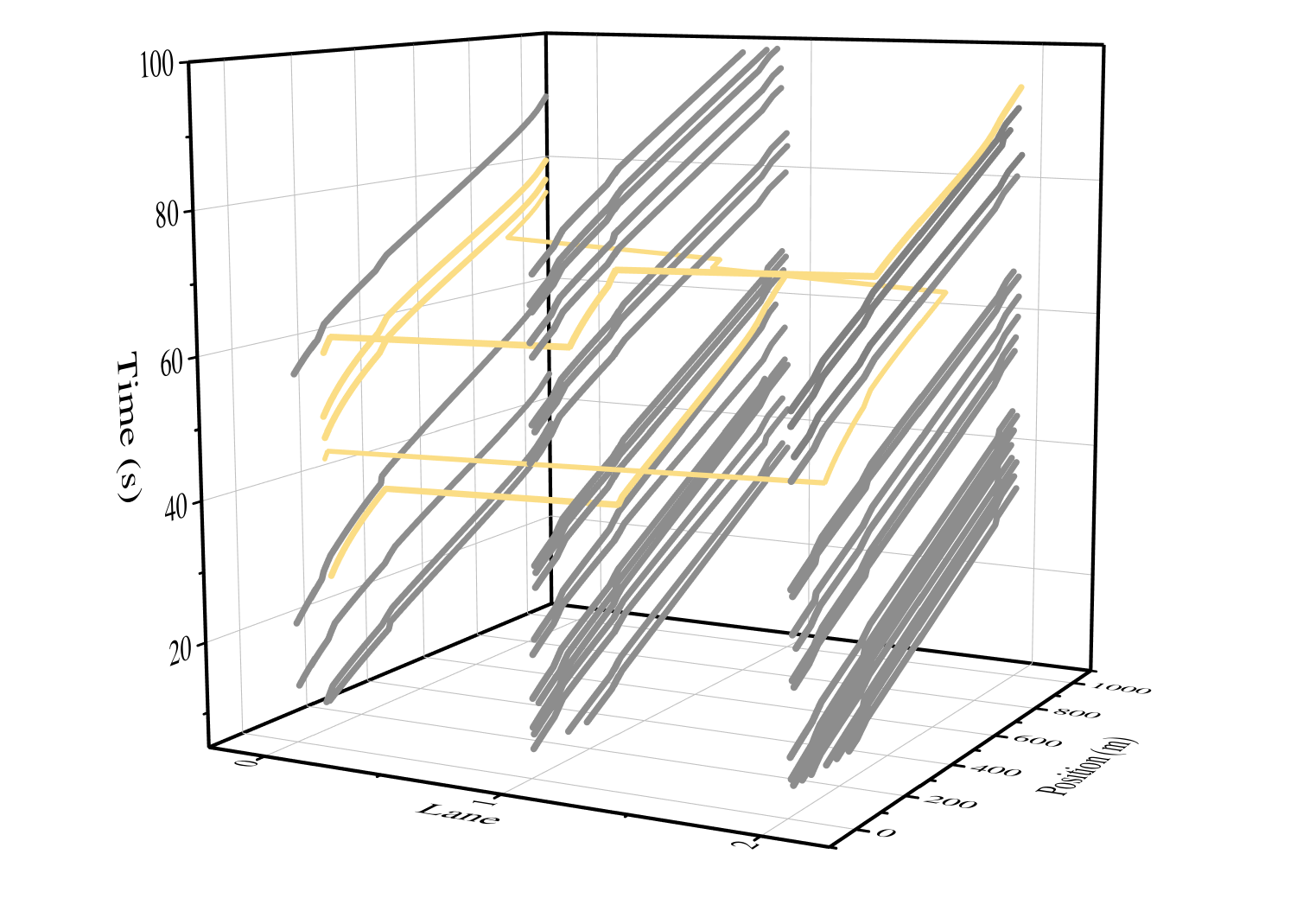}
            
        \end{minipage}
    }
    \subfigure[3D spatiotemporal trajectory test diagram of PAH-DNN.]{
        \begin{minipage}[t]{0.315\linewidth}
            \includegraphics[width=5.5cm]{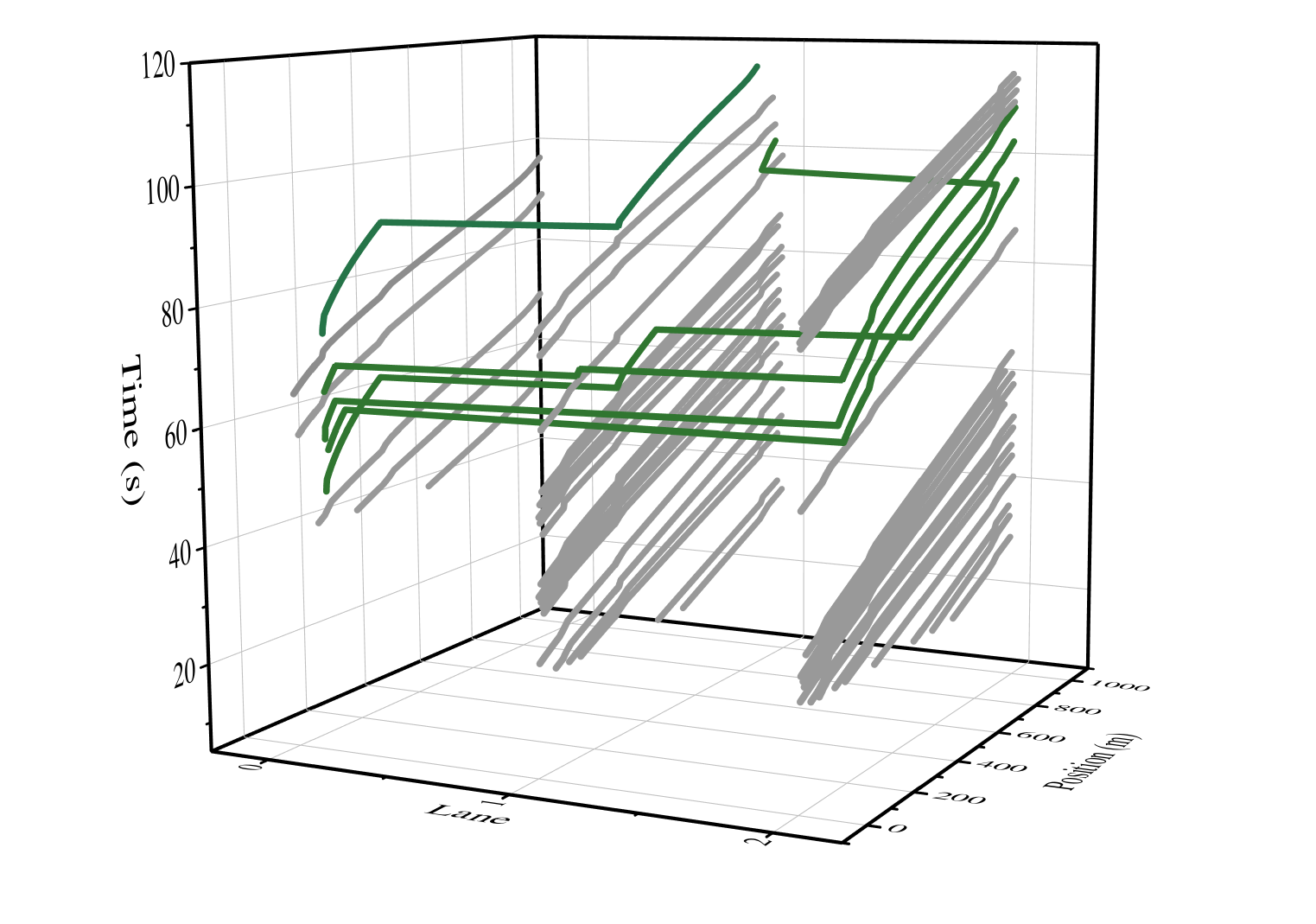}
            
        \end{minipage}
    }
    \subfigure[Lane-based spatiotemporal trajectory test diagram of PAH-MLGAT.]{
        \begin{minipage}[t]{0.315\linewidth}
            \includegraphics[width=5.5cm]{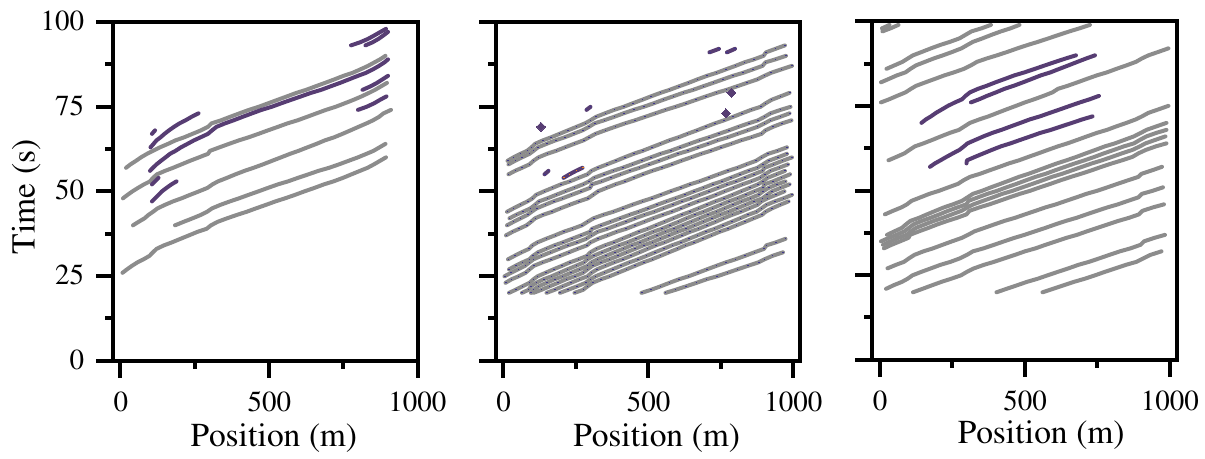}
            
        \end{minipage}
    } 
    \subfigure[Lane-based spatiotemporal trajectory test diagram of PAH-GAT.]{
        \begin{minipage}[t]{0.315\linewidth}
            \includegraphics[width=5.5cm]{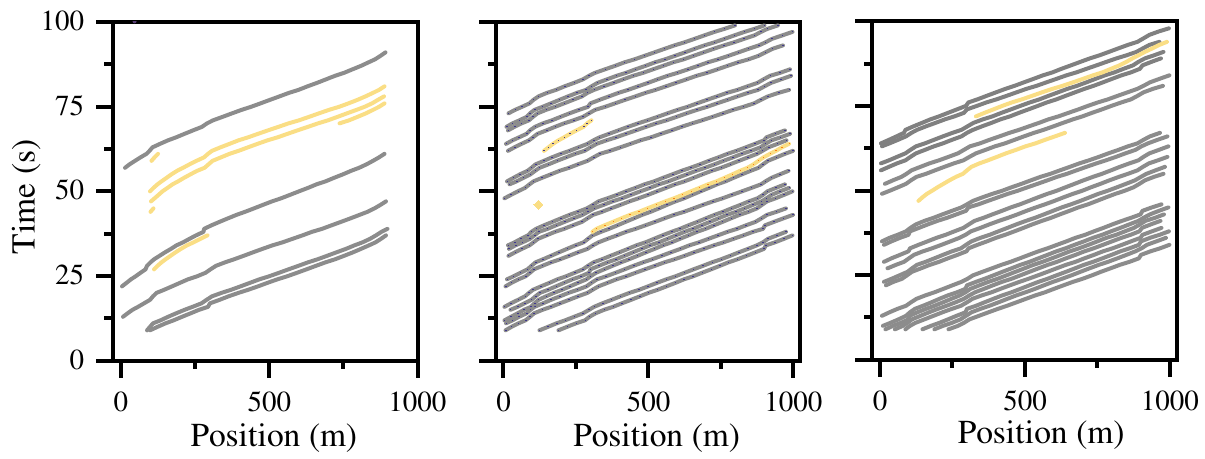}
            
        \end{minipage}
    }
    \subfigure[Lane-based spatiotemporal trajectory test diagram of PAH-DNN.]{
        \begin{minipage}[t]{0.315\linewidth}
            \includegraphics[width=5.5cm]{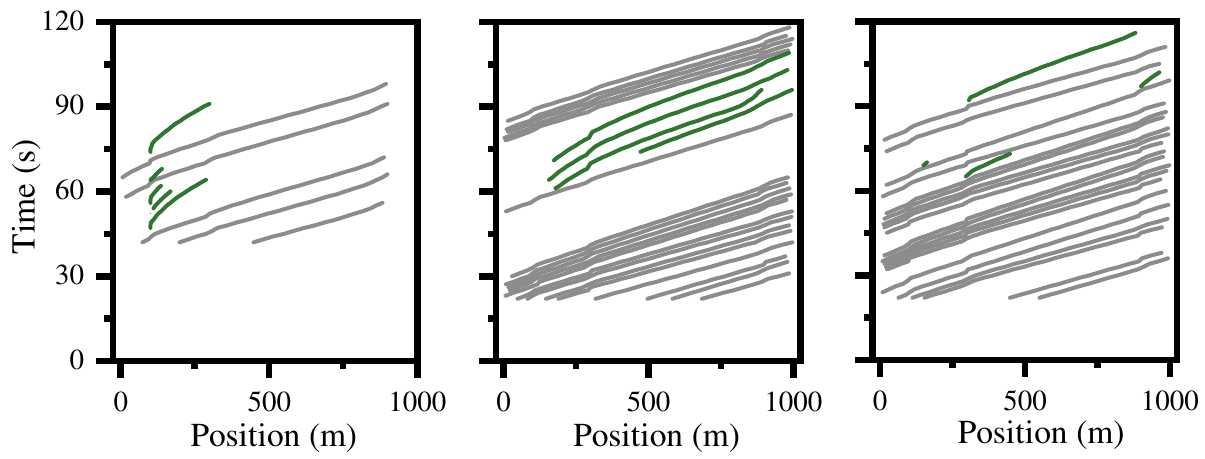}
            
        \end{minipage}
    }
    \caption{Spatiotemporal trajectory test diagram of five CAVs trained by  three  algorithms.   (a)-(c)  3D spatiotemporal trajectories; (d)-(f)  planar trajectories across lanes 0 to 2. Trajectories of Human-driven Vehicles (HVs) are   in gray, while those of CAVs trained by PAH-MLGAT, PAH-GAT, and PAH-DNN are respectively shown in blue, orange, and green.}
    \label{fig:tra}
\end{figure*}

\subsection{Analysis of   spatiotemporal trajectories}
 
% This study presents the spatio-temporal trajectory analysis for five CAVs trained via three comparative algorithms. Figures \ref{fig:tra}(a)-(c) illustrate the vehicles' 3D spatio-temporal trajectories, whereas Figures \ref{fig:tra}(d)-(f) each encompass three sub-figures, delineating the planar trajectories across lanes 0 to 2. Moreover, the trajectories of Human-driven Vehicles (HVs) are depicted in gray, while those of CAVs trained by PAH-MLGAT, PAH-GAT, and PAH-DNN are respectively shown in blue, orange, and green.

Utilizing the median of the test rewards as a basis, a set of spatiotemporal trajectory diagrams were selected to evaluate the decision-making   of PAH-MLGAT, PAH-GAT, and PAH-DNN. In Figure \ref{fig:tra}, the z-axis shows time; the x-axis represents lanes 0 to 2, corresponding to the slow, middle, and fast lanes; and the base plane maps out the spatial movement trajectories. A preliminary examination focuses on task completion rates. As depicted in Figure \ref{fig:tra}(a), CAVs trained via PAH-MLGAT predominantly succeeded in their tasks, despite the third CAV's failure to merge into the fast lane through traffic. This   indicates that PAH-MLGAT substantially explored all logical objectives, adeptly capturing the  spatiotemporal attributes of the traffic flow. In Figure \ref{fig:tra}(b), only three CAVs of PAH-GAT managed to complete their tasks, with only the second CAV successfully navigating the fast lane. This suggests that PAH-GAT identified objectives associated with utilizing the fast lane and ramp exit, but fell short in pinpointing suitable spatiotemporal openings in traffic. Figure \ref{fig:tra}(c) demonstrates that PAH-DNN's CAVs failed to fulfill their tasks, highlighting its inclination toward immediate fast-lane rewards without recognizing the rewards associated with distant destinations, thereby exhibiting a myopic aspect.

We conducted a trajectory analysis of CAVs before and after a lane change. In the 2D diagrams, a diminished slope signifies an ability to traverse longer distances in shorter time spans. As illustrated in Figure \ref{fig:tra}(d), upon entry from the ramp into the slow lane, the slope for PAH-MLGAT's CAVs lessens, signaling acceleration initiation. Notably, the snug trajectory alignment of the third CAV with the following vehicle indicates its mastery over the lane's optimal speed. The abbreviated trajectories of the remaining four CAVs in the middle lane imply precise acquisition of adjacent lane traffic flow information. As depicted in Figure \ref{fig:tra}(e), the first and fifth CAVs of PAH-GAT are ensnared between the traffic flows of the middle and fast lanes, and are prevented from changing lanes. This shows their challenges in balancing lane-changing and following decisions. Displayed in Figure \ref{fig:tra}(f), PAH-DNN's CAVs exhibit continuously increasing trajectory slopes, with a corresponding decrease in speed, signifying the model's inability to make viable following decisions post-stratification.

  PAH-MLGAT demonstrates an ability to undertake collaborative decisions for lane changing and speed adjustment in dense traffic, adeptly choosing unoccupied spatiotemporal regions. This corroborates the proposed framework's capacity for synchronized, human-like cognitive decision-making across both lane-changing and following dimensions.

\section{CONCLUSION}
To enable   CAVs to better capture spatiotemporal dependencies in non-Euclidean spaces within heterogeneous mixed autonomy, we introduced a decision-making strategy   with spatiotemporal interaction capabilities, which is designed to mimic human abilities for divided attention and cognitive consistence. Central to this approach is the multilevel graph representation theory, which addresses dynamic spatial correlations and nonlinear temporal interactions peculiar to non-Euclidean environments. We proposed a parallel asynchronous hierarchical graph reinforcement learning framework to enhance the autonomous learning and iterative refinement of CAVs, facilitating simultaneous strategy development across multiple dimensions despite limited attention capacities. This promotes cross-dimensional cooperation among CAVs, leading to decision-making processes that resemble human cognitive functions. To validate   our approach, we conducted   ablation studies and comparative analyses in a simulated heterogeneous mixed autonomy environment, emphasizing spatiotemporal interactions. Simulation results indicated that our PAH-MLGAT, integrating multilevel graph representation with PAH-GRL and multi-head graph attention mechanisms, substantially enhances  the decision-making performance and robustness of CAVs. Compared to PAH-DNN, our approach resulted in a sixfold increase in task success rate, a 77.70\% reduction in braking instances, an 18.57\% decrease in driving time, a 17.36\% reduction in energy consumption, and a 39.74\% reduction in average variance across six metrics.

For future research, it is suggested to refine the study of the hierarchical decision-making framework. Moreover,  it is imperative to further explore graph network models, particularly those based on traffic graph representations. From an experimental perspective, the development of a micro-traffic sandbox environment is advocated, which would enable hardware-in-the-loop experiments.
\appendices
\section{Proof of   Degree Distribution Imbalance}

 A graph at time-step \(t\) is considered as \(G_t = (V_t, E_t)\), where the disappearance of node \(u\) leads to the removal of all its connected edges. The adjacency matrix \(A_t\) represents graph \(G_t\), where \(a_{ij}^t = 1\) if nodes \(i\) and   \(j\) are connected at time \(t\), and is 0 otherwise.

The degree of node \(u\),  \(d_t(u)\), is defined as the number of elements with a value of 1 in the row (or column, as the graph is undirected) corresponding to node \(u\) in the adjacency matrix,
\begin{equation}
d_t(u) = \sum_{j \in V_t} a_{uj}^t.
\end{equation}    

At time \(t+1\), node \(u\) and its associated edges are removed, and hence \(a_{uj}^{t+1} = 0\) and \(a_{ju}^{t+1} = 0\) for all \(j\). The degree of any node \(v \neq u\) that was a neighbor of \(u\) becomes
\begin{equation}
d_{t+1}(v) = d_t(v) - 1.
\end{equation}  

 The degrees of all nodes directly connected to \(u\) decrease by 1, leading to a change in the degree distribution. In particular, the total number of edges \(|E_{t+1}| = |E_t| - d_t(u)\).

The degree distribution can be represented through a degree sequence or degree distribution function. Let \(P(k; t)\) be the proportion of nodes with degree \(k\) in the graph at time \(t\). After the disappearance of node \(u\), the proportions of nodes with degrees \(k\) and \(k-1\) will change. Specifically, for each node connected to \(u\),
\begin{equation}
P(k; t+1) = P(k; t) - \Delta P(k; t),
\end{equation}  
where \( \Delta P(k; t) \) is the decrease in the proportion of nodes with degree \(k\) due to node disappearance, and the proportion of nodes with degree \(k-1\) will increase accordingly:
\begin{equation}
P(k-1; t+1) = P(k-1; t) + \Delta P(k; t). 
\end{equation} 
Since this change is concentrated on nodes directly connected to \(u\), it causes a sudden and significant shift in the graph's degree distribution, causing an imbalance.

\section{Proof of   Theorem 1.}
Define the weighted adjacency matrix \(A_t \in [0,1]^{N \times N}\), where each element \(a_{ij}^t\) represents the interaction strength between nodes \(i\) and   \(j\) at time \(t\), which can be based on spatial distance and node features, as
\begin{equation}
 a_{ij}^t = f(\Delta d_{ij}, \mathbf{x}_i, \mathbf{x}_j),
\end{equation} 
where \(\Delta d_{ij}\) is the spatial distance between nodes \(i\) and \(j\), \(\mathbf{x}_i\) and \(\mathbf{x}_j\) are their respective feature vectors, and \(f\) is a mapping function that translates these inputs into a weight value in the interval [0,1].

In a weighted graph, the weighted degree of node \(i\) can be defined as
\begin{equation}
 d_t^{w}(i) = \sum_{j \in V_t} a_{ij}^t,
\end{equation} 
reflecting the total interaction strength of node \(i\) with other nodes.

When node \(u\) disappears, for any remaining node \(v\), the change in its weighted degree is no longer a simple subtraction of 1, but depends on the value of \(a_{uv}^t\),
\begin{equation}
d_{t+1}^{w}(v) = d_t^{w}(v) - a_{uv}^t.
\end{equation}
The weighted degrees of all nodes not directly connected to \(u\)  remain unchanged.

Since the weight \(a_{ij}^t\) reflects the combined impact of spatial distance and node features, as shown in equations \ref{weighted_L} and \ref{weighted_F}, when the distance threshold \((X,Y) \to +\infty\),
\begin{equation}
\lim_{\Delta{d_{ij}\to +\infty}}  f(\Delta d_{ij}, \mathbf{x}_i, \mathbf{x}_j) = 0,
\end{equation}
which implies that even with node disappearance, the degree distribution of the graph can adjust more smoothly, reducing abrupt changes.

% \section{}

% \section*{Acknowledgment}


\ifCLASSOPTIONcaptionsoff
  \newpage
\fi

% You can use other form of bib file by changing here...
\bibliographystyle{IEEEtran}
% \bibliography{reference.bib}
Generated by IEEEtran.bst, version: 1.14 (2015/08/26)

\end{document}